# High solubility of water in post-perovskite and bridgmanite in the Earth's deep lower mantle

Megumi Semba[1], Naoya Sakamoto[2], Shuhei Mita[1], Kei Hirose[1,3], Yutaro Tsutsumi[1], Keisuke Ozawa[4], John W. Hernlund[3] & Shiro Miwa[2]

[1]Department of Earth and Planetary Science, The University of Tokyo, Bunkyo, Tokyo 113-0033, Japan

[2]Institute for Integrated Innovations, Hokkaido University, Sapporo, Hokkaido 001-0021, Japan

[3]Earth-Life Science Institute, Institute of Science Tokyo, Meguro, Tokyo 152-8550, Japan

[4]Graduate School of Science, University of Hyogo, Kamigori-cho, Ako-gun, Hyogo 678-1297, Japan

**Water in the Earth's mantle induces melting, giving rise to strong chemical heterogeneity, and alters its rheological and transport properties, which are keys to understanding seismic-wave speeds anomalies, convective motions and electrical conductivity. However, the abundance of water and its role in the deep mantle remain uncertain and controversial, particularly at conditions of the core-mantle boundary (CMB) region. Here we carry out melting experiments on a hydrous, natural mantle composition including both $H_2O$ and $D_2O$ over the entire pressure range of the Earth's lower mantle, in which melt coexists with bridgmanite (Bdg) and/or post-perovskite (PPv), the primary constituents of the respective lower and lowermost mantle. High-resolution secondary-ion mass spectroscopy (Cryo-SIMS) measurements reveal high water concentrations in Bdg (up to 5,500 ppm by weight) and PPv (up to 1.22 wt%) with strong enrichment in deuterium/hydrogen (D/H) relative to coexisting melt. The high solubilities of water in Bdg and PPv at CMB conditions suggest that subducting slabs do not release water when they reach the bottom of the mantle, and such a process cannot account for strong chemical heterogeneities inferred in the lowermost mantle[1–3] and the topmost outer core[4], nor for ultralow-velocity zone (ULVZ)-like structures in seismically fast regions[5,6]. Water-rich Bdg and PPv may be present in the large low shear velocity provinces and ULVZs, which may be the residue of a basal magma ocean and are predicted to host a deep low D/H water reservoir inherited from the early Earth that is occasionally sampled by plumes of a lowermost-mantle origin[7,8].**

Even at small concentrations, water strongly influences the properties of the rocky mantle, and both the solubility and actual amount of $H_2O$ residing within major mantle minerals have been extensively studied by experiment and theory[9]. However, experiments have been limited to ~70 GPa[10], corresponding to mid-mantle depths, although the distribution of water and its effect on seismic and transport properties at deeper conditions have been examined by theory[11–14]. Furthermore, the capacity of water in Bdg has long been highly controversial, with previous experimental reports ranging from ~10 ppm[15,16] to ~1,000 ppm[17–23]. It has been argued[16] that reports of higher water abundance in Bdg is caused by inclusions of melt or superhydrous phase B captured within Bdg grains and/or at grain boundaries[24]. A recent experimental study[10] reported up to ~2,000 ppm $H_2O$ in Bdg at

~40 GPa, which did not include any nanometre-scale hydrous inclusions as evidenced by atom probe tomography.

We conducted new melting experiments on ($H_2O$, $D_2O$)-bearing pyrolitic mantle material to examine the water solubility in Bdg and PPv and the partitioning of water between Bdg/PPv and melt at shallow to deep lower mantle pressures. In a portion of the present experiments, after sample melting at high pressure, all processes—from sample decompression to SIMS analyses—were performed under cryogenic temperatures[25] without sample exposure to air. The difference in the D/H ratio between Bdg and melt confirms that water found in Bdg does not derive from melt inclusions nor melt at grain boundaries, but rather resides in its crystal lattice. The SIMS measurements reveal significantly higher amounts of water in primary lower mantle crystals of Bdg and PPv than previously reported in experiments, even at high temperatures coexisting with melts, which carries profound implications for the behaviour and abundance of water in the Earth's deep mantle. The observed isotopic fractionation of hydrogen between Bdg/PPv and melt also provides an alternative mechanism to explain low D/H ratios found in magmas originating from deep-rooted plumes[7,8].

**Solubility, distribution and isotopic fractionation of water in the lower mantle**

Thirteen separate experiments were performed in the laser-heated diamond-anvil cell (DAC) across a wide pressure and temperature (*P*-*T*) range (36–141 GPa and 3,340–4,860 K) (see Methods, Extended Data Fig. 1a). After melting the hydrous pyrolite sample, the liquidus phase coexisting with quenched melt (Bdg and/or PPv) was confirmed by synchrotron X-ray diffraction (XRD) measurements (Extended Data Fig. 2). In half of the present experiments, the sample was then decompressed and recovered from the DAC in liquid $N_2$, and was transferred to, and also analysed by, a SIMS instrument (CAMECA, ims-1270e7) under continuous cryogenic temperatures[25] without exposure of the sample to air. Such procedures minimize potential hydrogen migration after the sample was melted under high pressure and ensure that the sample is not contaminated by hydrogen in the ambient air (Extended Data Table 1). In other runs, we quickly released pressure, recovered the sample in a $N_2$ atmosphere, and transferred it to the Cryo-SIMS instrument, also without exposing the sample to air. In both cases, the sample on a liquid $N_2$-cooled SIMS stage was pre-sputtered until a quenched melt pool was exposed on the sample surface (Fig. 1, Extended Data Fig. 3). The round melt pool was surrounded by a single-phase layer of Bdg or PPv in all samples, which was confirmed by textural and compositional characterizations using electron microprobes. In run #9 (only) our microprobe observations revealed a PPv layer outside the Bdg layer (Fig. 1b), consistent with XRD data.

Two-dimensional SIMS images of $^{1}H^{+}$ and $^{2}D^{+}$ demonstrate that water is enriched in the central melt pool but is also present in the Bdg/PPv layer (Fig. 1, Extended Data Table 1). The quantified water contents ($H_2O$ + $D_2O$ combined) in Bdg ranged from 1,850(130) ppm at 41 GPa/3,920 K to 5,510(380) ppm at 77 GPa/3,930 K (Fig. 2a, Extended Data Fig. 4a, see Methods). While the water capacity of Bdg has been highly controversial[15,16], a recent single-crystal neutron diffraction study[18] determined the crystallographic site position of hydrogen in Bdg, and argued for a capacity of ~1,000 ppm $H_2O$ at 24 GPa. With ultrahigh-resolution atom probe tomography demonstrating the absence of nm-scale inclusions of melt and other water-bearing phases within Bdg grains, up to ~2,000 ppm $H_2O$ was reported at 40 GPa[10]. Our findings of 1,850–5,510 ppm water in Bdg at 36–117

GPa are consistent with these recent experimental reports of ~1,000–2,000 ppm $H_2O$ at relatively low *P*-*T* ranges and much higher under deep lower-mantle conditions than previously reported in the literature[19,22–24]. Furthermore, we found that PPv contained higher amounts of water than these Bdg crystals. The SIMS analyses revealed 0.652(17) to 1.223(26) wt% $H_2O$ + $D_2O$ in PPv formed at 117–141 GPa/3,870–4,860 K (Figs. 1b, 2a, Extended Data Fig. 4a).

The amount of water in Bdg/PPv was controlled by partitioning between these crystals and melts, in addition to the amount of water involved in melting which is sensitive to the laser-heated volume in excess of the solidus temperature. The present experiments conducted over a wide pressure range demonstrate that the Bdg/melt partition coefficient of $H_2O$, $D_{H2O}$(Bdg/melt) ranged from 0.024(3) to 0.113(12) by weight (Extended Data Table 1), increasing at higher pressures and temperatures. We newly find a positive pressure-dependence for water solubility at pressures higher than prior experiments[10,19] (Fig. 2b):

$$\log_{10} D_{H2O}(\text{Bdg/melt}) = -0.06(44) - 6.2(14) \times 10^3/T + 20.6(87) \times P/T \qquad (1)$$

where $T$ is temperature in kelvin and $P$ is pressure in gigapascals. We note that the $D_{H2O}$(Bdg/melt) data measured in this study and by Lu *et al.*[10] are well fitted by Eq. 1 (Fig. 2b), indicating that Bdg was not saturated with $H_2O$ in either study except for our run #4 (once saturated, $D_{H2O}$ apparently diminishes, Extended Data Fig. 5) and the maximum solubility of water can be higher than observed. Water is distributed substantially more into PPv than to Bdg, exhibiting $D_{H2O}$(PPv/melt) = 0.112(13) to 0.281(31). Bdg and PPv were in direct contact in run #9, and the PPv hosted twice the amount of water as the Bdg (Fig. 1b):

$$D_{H2O}(\text{PPv/melt}) = 2 \times D_{H2O}(\text{Bdg/melt}) \qquad (2)$$

The major element compositions of Bdg and PPv (Extended Data Table 1) indicate that the charge-coupled substitution $Si^{4+} = Al^{3+} + H^+$ is an important mechanism for water incorporation into these crystals (Extended Data Fig. 6). It is consistent with recent single-crystal neutron diffraction measurements of Bdg[18], while other mechanisms such as $Mg^{2+} = 2H^+$ and $Si^{4+} = 4H^+$ were proposed for Bdg[20]. The $Si^{4+} = 4H^+$ substitution could be an additional water incorporation mechanism into PPv[12] since 1) PPv included more $H^+$ + $D^+$ than $Al^{3+}$ and 2) PPv exhibited Si depletions (Extended Data Figs. 6a, c), which are not reconciled with the $Si^{4+} = Al^{3+} + H^+$ substitution but suggest the Si vacancy formation (the $Si^{4+} = Al^{3+} + H^+$ substitution causes significant volume expansion[26]), a part of which is occupied by $4H^+$ (Extended Data Figs. 6b, d).

The present SIMS measurements with fully cryogenic sample processing[25] also reveal a remarkable difference in D/H ratio between coexisting melt and Bdg (Extended Data Table 1), more than can be explained by the matrix effect in SIMS analysis[27,28] (see Methods). This indicates significant hydrogen isotopic fractionation upon melting/crystallization in the lower mantle, leading to the enrichment of hydrogen relative to deuterium in partial/residual melt. The observed isotopic fractionation factor, $\alpha = (D/H)_{Bdg}/(D/H)_{melt}$ ranged from 1.15(5) at 77 GPa/3,930 K to 1.30(6) at 36 GPa/3,340 K (Fig. 3). The isotopic fractionation factor is higher at lower temperatures, following[29,30]:

$$\alpha = 1 + a/T^2 \qquad (3)$$

Our data give $a = 2.95(26) \times 10^6$, suggesting that noticeable hydrogen isotopic fractionation could occur even at a liquidus temperature of around 4700–5300 K[31,32] in the CMB region. In addition, we note that the D/H ratio of PPv was 0.299(40) in run #9, which matches 0.316(35) for coexisting Bdg within our uncertainties.

**Fate of water transported by subducting slabs into the deep lower mantle**

These results demonstrate that PPv is capable of hosting more than 1 wt% water under the Earth's CMB conditions. Bdg can accommodate >5,000 ppm water at 77 GPa and 3,930 K, and likely more in the lowermost mantle because the pressure effect on water solubility is expected to be positive (note that these water concentrations far exceed the capacity of water in Bdg at ~25 GPa and ~2,000 K of about 1,000 ppm[17–21] regardless of the negative temperature effect[33]). In addition, while both Bdg and PPv in our experiments included 2–3 wt% $Al_2O_3$ (Extended Data Table 1), the water capacity of crystals containing ~5 wt% $Al_2O_3$ in a pyrolitic mantle is expected to be even greater since $Si^{4+} = Al^{3+} + H^+$ substitution is the primary mechanism of water incorporation, at least at high temperatures (>3,300 K) examined in the present experiments.

The high capacity of water in Bdg, and more so in PPv, at conditions of Earth's deep mantle carries important implications for the structure and evolution of the CMB region. One immediate inference of these results is that water release via dehydration of Bdg and PPv is unlikely to occur in deeply subducted lithospheric slabs at the base of the mantle because significantly less $H_2O$ is likely to reach the deepest mantle than the solubility limits of these phases. While hydrous phases may carry water in cold subducting slabs down to the mid-lower mantle[34], they decompose at 60 GPa and 1,500 K[1], releasing water to surrounding nominally anhydrous minerals, mainly Bdg in the slab peridotite layer and possibly the $SiO_2$ phase if crustal materials are present[34]. Assuming 100% $H_2O$ as a melt phase as an upper bound, $H_2O$ incorporation into Bdg will be less than ~1,000 ppm for $D_{H2O}$(Bdg/melt) = 0.0004 at corresponding conditions (Eq. 1). Once the cold slabs reach the lowermost mantle, Bdg undergoes a phase transformation into PPv. The PPv in subducted slabs is Al-poor (<1 wt%)[35] because hydration of the slab mantle layer extends only to the uppermost depleted harzburgitic part, while the PPv formed in our experiments includes ~3 wt% $Al_2O_3$. The amount ~1,000 ppm $H_2O$ transported to the deep mantle from shallower depths is far less than the solubility of water in Al-poor PPv at the CMB, even when the Al content is proportional to the capacity of water. It is possible that the PPv transforms back to Bdg under high temperatures of the CMB region[36] (see Methods), although the depletion in Al favours PPv[37] and the enrichment of water in PPv relative to Bdg by a factor of two also indicates that water stabilizes PPv[12]. Even so, water-bearing Al-poor Bdg will not dehydrate at the CMB because its water capacity is higher than 5,000 ppm with 2.8 wt% $Al_2O_3$ and should be as high as ~2,000 ppm with ~1 wt% $Al_2O_3$ corresponding to $Al^{3+} : H^+ = 1 : 1$ that maximizes the $Si^{4+} = Al^{3+} + H^+$ substitution. Retention of ample water in the $SiO_2$ phase included in subducted crust at the CMB has also been reported[38]. Therefore, slabs entering the CMB region bearing ~1,000 ppm water in PPv or Bdg will not release it into the surrounding mantle, and this water will be returned to the shallower mantle.

Partial melting in the lowermost mantle induced by dehydration of deeply subducted lithosphere has been invoked to explain the presence of strong seismic low velocity anomalies in the vicinity of higher seismic velocity regions interpreted to be ponded slabs[5,6]. Expelled slab water has also been proposed to react with the topmost outer core to produce a stably stratified layer[4]. The present experiments, however, suggest that such scenarios are unlikely. Ongoing slab dehydration-induced partial melting is incompatible with the lack of seismic evidence for runaway accumulation of melts that would be expected to occur in the CMB region over geological time[39], a problem that is exacerbated by the requisite degree of CMB secular cooling associated with sustenance of a geodynamo for billions of years[40]. Such considerations motivate an alternative scenario that explains structures such as ULVZs and large low shear velocity provinces (LLSVPs) as the result of freezing rather than ongoing melting processes[41]. This "basal magma ocean" (BMO) model provides an alternative framework for interpreting the structures and evolution of the Earth's deep interior, and strongly influences the fate of water in the CMB region.

**Crystallization of hydrous Bdg and PPv in a basal magma ocean**

Our present results open new possibilities for the BMO and implications for LLSVPs. The BMO would become enriched in water by fractionation, since water still prefers melt to Bdg and/or PPv under lowermost mantle conditions (Fig. 2b). This will result in an increase in water concentration in the CMB region as fractional crystallization proceeds. Because both ferrous iron and water stabilize PPv, evolved cumulates of the BMO may include hydrous iron-rich PPv, which could then constitute a portion of the LLSVPs. Indeed, co-existing Bdg+PPv in run #9 at 117 GPa suggests that the triple point of Bdg-PPv-melt at $H_2O$-rich conditions is significantly lower than the CMB pressure (136 GPa), providing a strong argument for the stability of hydrous PPv in the lowermost mantle even at the highest possible temperatures. Previous studies[42] suggested the presence of PPv-like seismic discontinuities inside the Pacific LLSVP based on the possibility that iron could stabilize PPv even at the high temperatures expected in chemically distinct piles persisting over long time scales at the CMB. The fact that water also stabilizes PPv increases the likelihood of this scenario. Enrichment in both FeO and $H_2O$ in PPv causes a strong reduction in the shear-wave velocity with relatively modest changes in the bulk sound velocity[12], compatible with seismological observations of the LLSVPs[43]. Moreover, residual BMO melt would become progressively lower in D/H as both Bdg and PPv prefer deuterium over hydrogen (Fig. 3).

We model the evolution of water concentration and hydrogen isotopic composition in the BMO (Fig. 4), based on a recent BMO solidification scenario[44] including the crystallization of PPv in the lowermost mantle (see Methods). As BMO solidification proceeds, crystallizing solids eventually become dense enough to remain stable against complete entrainment by mantle convection, possibly when the amount of residual melt reduces to a size similar to the observed LLSVPs (~2% of the entire mantle)[45]. Since then, the BMO may have further evolved to the present-day ULVZs while crystallizing hydrous PPv to grow dense LLSVPs (Fig. 4c). LLSVP and ULVZ materials may not be straightforwardly sampled at Earth's surface, and if they could be fully isolated from a convecting mantle, our simulations suggest that they could maintain an important water reservoir, sequestering 35% of $H_2O$ and 33% of $D_2O$ that were originally in a fully-molten mantle magma ocean prior to crystallization (after core-forming metal segregation) (Fig. 4a). Its complement consists of accessible reservoirs—convecting mantle combined with lithospheric mantle, crust and oceans, which exhibit average $H_2O$ concentrations of 710 ppm[46] and hydrogen isotopic composition of $\delta D = -38‰$ (see Methods). With these observational constraints for the complement, our model gives 1.9 wt% $H_2O$ and exceedingly low D/H ratio ($\delta D = -124‰$) for sequestered materials on average (Figs. 4b, d). We also found far greater water concentrations and lower D/H values for ULVZs. The PPv crystals, which were originally formed in this water-rich BMO and are now present in the deeper portions of LLSVPs and mushy ULVZs, may include >0.4 wt% and >1.0 wt% $H_2O$, respectively (Fig. 4c). The shallower portions of LLSVPs would be in the hydrous Bdg stability field at present, possibly holding up to ~1 wt% $H_2O$ with 5 wt% $Al_2O_3$. In this scenario a Bdg-PPv phase change would persist at the mid-levels of LLSVPs as they are stirred by the convective motions in the surrounding mantle, potentially producing seismic discontinuities[42]. Hydrous PPv and Bdg cumulates in the LLSVPs will possess positive δD values (Fig. 4d), but if entrained in upwelling plumes along with BMO residuum, they could release water exhibiting exceedingly low D/H ratios (as low as δD = –400 to –510 ‰ when α = 1.7 at 2,000 K, Fig. 3) at middle to shallow lower-mantle depths as Bdg water capacity diminishes. Such D-depleted water might explain strongly negative δD inclusions found in Hawaiian Koolau basalts[8] and proto-Iceland Baffin Island picrites with δD as low as –218‰[7].

Finally, we note that the Earth's magma ocean (assuming a fully molten mantle) may have included 1,075 ppm $H_2O$, when combining water in the present-day LLSVP+ULVZ and its complementary reservoirs (see Methods). Core-mantle distributions would then suggest core hydrogen concentrations of at least 0.3 wt%, higher than earlier estimates[47,48] that were based on metal-silicate partitioning employing a lower value of 710 ppm $H_2O$[46]. Core-forming metals underwent hydrogen isotope exchange with silicate melt in a magma ocean that possessed δD = –68‰ (see Methods), which carries important implications for the D/H ratios of the core and the bulk Earth once the hydrogen isotope fractionation between metal and silicates can be clarified.

1. Ohtani, E. The role of water in Earth's mantle. *Natl. Sci. Rev.* **7**, 224–232 (2020).
2. Hu, Q. *et al.* $FeO_2$ and FeOOH under deep lower-mantle conditions and Earth's oxygen–hydrogen cycles. *Nature* **534**, 241–244 (2016).
3. Liu, J. *et al.* Hydrogen-bearing iron peroxide and the origin of ultralow-velocity zones. *Nature* **551**, 494–497 (2017).
4. Kim, T. *et al.* A hydrogen-enriched layer in the topmost outer core sourced from deeply subducted water. *Nat. Geosci.* **16**, 1208–1214 (2023).
5. Hansen, S. E., Garnero, E. J., Li, M., Shim, S.-H. & Rost, S. Globally distributed subducted materials along the Earth's core-mantle boundary: Implications for ultralow velocity zones. *Sci. Adv.* **9**, eadd4838 (2023).
6. Thorne, M. S., Takeuchi, N. & Shiomi, K. Melting at the edge of a slab in the deepest mantle. *Geophys. Res. Lett.* **46**, 8000–8008 (2019).
7. Hallis, L. J. *et al.* Evidence for primordial water in Earth's deep mantle. *Science* **350**, 795–797 (2015).
8. Hauri, E. SIMS analysis of volatiles in silicate glasses, 2: Isotopes and abundances in Hawaiian melt inclusions. *Chem. Geol.* **183**, 115–141 (2002).
9. Peslier, A. H., Schönbächler, M., Busemann, H. & Karato, S.-I. Water in the Earth's interior: Distribution and origin. *Space Sci. Rev.* **212**, 743–810 (2017).
10. Lu, W. *et al.* Substantial water retained early in Earth's deep mantle. *Science* **390**, 1177–1180 (2025).
11. Hernández, E. R., Alfè, D. & Brodholt, J. The incorporation of water into lower-mantle perovskites: A first-principles study. *Earth Planet. Sci. Lett.* **364**, 37–43

(2013).
12. Jiang, J. & Zhang, F. Theoretical studies on the hydrous lower mantle and D″ layer minerals. *Earth Planet. Sci. Lett.* **525**, 115753 (2019).
13. Mohn, C. E., Caracas, R. & Conrad, C. P. Lower mantle water distribution from ab initio proton diffusivity in bridgmanite. *Earth Planet. Sci. Lett.* **649**, 119095 (2025).
14. Peng, Y. & Deng, J. Hydrogen diffusion in the lower mantle revealed by machine learning potentials. *J. Geophys. Res. Solid Earth* **129**, e2023JB028333 (2024).
15. Bolfan-Casanova, N., Keppler, H. & Rubie, D. C. Water partitioning at 660 km depth and evidence for very low water solubility in magnesium silicate perovskite. *Geophys. Res. Lett.* **30**, 2003GL017182 (2003).
16. Liu, Z. *et al.* Bridgmanite is nearly dry at the top of the lower mantle. *Earth Planet. Sci. Lett.* **570**, 117088 (2021).
17. Fu, S. *et al.* Water concentration in single-crystal (Al,Fe)-bearing bridgmanite grown from the hydrous melt: Implications for dehydration melting at the topmost lower mantle. *Geophys. Res. Lett.* **46**, 10346–10357 (2019).
18. Purevjav, N. *et al.* Hydrogen incorporation mechanism in the lower-mantle bridgmanite. *Am. Mineral.* **109**, 1036–1044 (2024).
19. Xie, L. *et al.* Crystallization of a hydrous magma ocean in the shallow lower mantle. *Earth Planet. Sci. Lett.* **633**, 118651 (2024).
20. Zhang, Y. *et al.* Hydrogen dissolution mechanisms in bridgmanite by first-principles calculations and infrared spectroscopy. *J. Geophys. Res. Solid Earth* **130**, e2024JB030403 (2025).
21. Inoue, T., Wada, T., Sasaki, R. & Yurimoto, H. Water partitioning in the Earth's mantle. *Phys. Earth Planet. Inter.* **183**, 245–251 (2010).
22. Litasov, K. *et al.* Water solubility in Mg-perovskites and water storage capacity in the lower mantle. *Earth Planet. Sci. Lett.* **211**, 189–203 (2003).
23. Murakami, M., Hirose, K., Yurimoto, H., Nakashima, S. & Takafuji, N. Water in Earth's lower mantle. *Science* **295**, 1885–1887 (2002).
24. Amulele, G., Karato, S. & Girard, J. Melting of bridgmanite under hydrous shallow lower mantle conditions. *J. Geophys. Res. Solid Earth* **126**, e2021JB022222 (2021).
25. Sakamoto, N., Ikuta, N., Hirose, K., Hikosaka, K. & Mita, S. SIMS preserving elemental distributions formed under extreme conditions: Hydrogen distribution measurement in iron. *Surf. Interface Anal.* https://doi.org/10.1002/sia.70091 (2026).
26. Tsuchiya, J., Tsuchiya, T. & Wentzcovitch, R. M. Transition from the $Rh_2O_3$(II)-to-$CaIrO_3$ structure and the high-pressure-temperature phase diagram of alumina. *Phys. Rev. B* **72**, 020103 (2005).
27. Hauri, E. H. *et al.* Matrix effects in hydrogen isotope analysis of silicate glasses by SIMS. *Chem. Geol.* **235**, 352–365 (2006).
28. Sobolev, A. V. *et al.* Deep hydrous mantle reservoir provides evidence for crustal recycling before 3.3 billion years ago. *Nature* **571**, 555–559 (2019).
29. Bigeleisen, J. & Mayer, M. G. Calculation of equilibrium constants for isotopic exchange reactions. *J. Chem. Phys.* **15**, 261–267 (1947).
30. Young, E. D. *et al.* High-temperature equilibrium isotope fractionation of non-traditional stable isotopes: Experiments, theory, and applications. *Chem. Geol.* **395**, 176–195 (2015).
31. Andrault, D. *et al.* Solidus and liquidus profiles of chondritic mantle: Implication for melting of the Earth across its history. *Earth Planet. Sci. Lett.* **304**, 251–259

(2011).
32. Fiquet, G. *et al.* Melting of peridotite to 140 gigapascals. *Science* **329**, 1516–1518 (2010).
33. Dong, J., Fischer, R. A., Stixrude, L. P. & Lithgow-Bertelloni, C. R. Constraining the volume of Earth's early oceans with a temperature-dependent mantle water storage capacity model. *AGU Adv.* **2**, e2020AV000323 (2021).
34. Walter, M. J. Water transport to the core–mantle boundary. *Natl. Sci. Rev.* **8**, nwab007 (2021).
35. Michael, P. J. & Bonatti, E. Peridotite composition from the North Atlantic: Regional and tectonic variations and implications for partial melting. *Earth Planet. Sci. Lett.* **73**, 91–104 (1985).
36. Hernlund, J. W., Thomas, C. & Tackley, P. J. A doubling of the post-perovskite phase boundary and structure of the Earth's lowermost mantle. *Nature* **434**, 882–886 (2005).
37. Akber-Knutson, S., Steinle-Neumann, G. & Asimow, P. D. Effect of Al on the sharpness of the $MgSiO_3$ perovskite to post-perovskite phase transition. *Geophys. Res. Lett.* **32**, 2005GL023192 (2005).
38. Tsutsumi, Y. *et al.* Retention of water in subducted slabs under core–mantle boundary conditions. *Nat. Geosci.* **17**, 697–704 (2024).
39. Hernlund, J. W. & Tackley, P. J. Some dynamical consequences of partial melting in Earth's deep mantle. *Phys. Earth Planet. Inter.* **162**, 149–163 (2007).
40. Labrosse, S. Thermal evolution of the core with a high thermal conductivity. *Phys. Earth Planet. Inter.* **247**, 36–55 (2015).
41. Labrosse, S., Hernlund, J. W. & Coltice, N. A crystallizing dense magma ocean at the base of the Earth's mantle. *Nature* **450**, 866–869 (2007).
42. Lay, T., Hernlund, J., Garnero, E. J. & Thorne, M. S. A post-perovskite lens and *D"* heat flux beneath the Central Pacific. *Science* **314**, 1272–1276 (2006).
43. Hernlund, J. W. & Houser, C. On the statistical distribution of seismic velocities in Earth's deep mantle. *Earth Planet. Sci. Lett.* **265**, 423–437 (2008).
44. Ozawa, K. *et al.* Trace element partitioning in a deep magma ocean and the origin of the Hf-Nd mantle array. *Sci. Adv.* **10**, eadp0021 (2024).
45. Burke, K., Steinberger, B., Torsvik, T. H. & Smethurst, M. A. Plume generation zones at the margins of large low shear velocity provinces on the core–mantle boundary. *Earth Planet. Sci. Lett.* **265**, 49–60 (2008).
46. Hirschmann, M. M. Comparative deep Earth volatile cycles: The case for C recycling from exosphere/mantle fractionation of major ($H_2O$, C, N) volatiles and from $H_2O$/Ce, $CO_2$/Ba, and $CO_2$/Nb exosphere ratios. *Earth Planet. Sci. Lett.* **502**, 262–273 (2018).
47. Tagawa, S. *et al.* Experimental evidence for hydrogen incorporation into Earth's core. *Nat. Commun.* **12**, 2588 (2021).
48. Tsutsumi, Y. *et al.* Origin of Earth's hydrogen and carbon constrained by their core-mantle partitioning and bulk Earth abundance. *Nat. Commun.* **16**, 10038 (2025).

## Methods

### High *P-T* experiments

We performed melting experiments using a laser-heated DAC with flat 300 μm and bevelled 120 and 90 μm culet anvils, depending on a pressure of target. Starting material was a powder mixture of an MgO-depleted peridotite material and $Mg[O(H,D)]_2$ (Extended Data Table 1) with a chemical composition of peridotite KLB-1 (44.5% $SiO_2$, 0.2% $TiO_2$, 3.6% $Al_2O_3$, 8.1% FeO, 39.2% MgO, 3.4% CaO and 0.3% $Na_2O$ by weight without water)[49]. The peridotitic material was synthesized originally from gel and then dried and reduced in an $H_2$-$CO_2$ mixed gas furnace under the oxygen fugacity three log units below the QFM buffer. The SIMS analyses of an unheated portion of a recovered DAC sample indicate that the starting power mixture possessed locally heterogeneous water concentrations of 1.26±0.38 wt% $H_2O$ and 0.19±0.08 wt% $D_2O$ (Fig. 1a). After compression to a pressure of interest, the sample was heated from both sides with a couple of 100 W single-mode Yb fibre lasers. We used beam-shaping optics, which converts a Gaussian beam to one with a flat energy distribution and thus diminishes the radial temperature gradient in the sample. The laser-heated spot was ~20–40 μm across. One-dimensional temperature distributions on a sample were measured using the spectro-radiometric method[50] (Extended Data Fig. 7). Since chemical equilibrium including the partitioning of water was attained at the boundary between the melt pool and a solid Bdg (or PPv) layer, we employ the temperature at the boundary as the experimental temperature of each run, which was obtained from the one-dimensional temperature profile and the size of the melt pool (Extended Data Fig. 7). Such a method to estimate crystallization temperature is practically equivalent to that by Lu *et al.*[10]. Heating duration was 3–10 s in order to avoid temperature fluctuation that could lead to a complex melting texture. Earlier time-series experiments[38] found that 1 s is long enough for the attainment of equilibrium water partitioning between melt and the $SiO_2$ phase in a laser-heated DAC under high *P-T* conditions similar to the present ones. Pressure was determined from the Raman shift of the diamond anvil at 300 K after heating and corrected for thermal pressure of 2.5 GPa per 1000 K following refs. 10, 31 and 51. Pressure and temperature uncertainties may be ±10% and ±5%, respectively[38,52].

For runs #3, #6–9 and #11–13, XRD patterns were collected at high pressures in a DAC after temperature was quenched to 300 K at the beamline BL10XU, SPring-8 synchrotron radiation facility (Extended Data Fig. 2). We used a monochromatic X-ray beam with an energy of ~30 keV, which was focused to 6 μm area (full width at half maximum) on the sample position. These XRD observations elucidated the liquidus phase (Bdg and/or PPv) surrounding a quenched melt pool. As for runs #7 and #8, we acquired the XRD data also during heating and observed that no crystals appeared from melt when it was quenched to room temperature, indicating that the Bdg and PPv crystals observed after temperature quench were not quench crystals but rather the liquidus phases that coexisted with melt at high temperatures.

### SIMS analyses

We performed SIMS measurements at Hokkaido University to obtain the $H_2O$ and $D_2O$ contents in coexisting Bdg/PPv and melt and the hydrogen isotopic fractionation factor between them. In runs #1–5 and #9 (Extended Data Table 1), after melting a sample in the DAC, all procedures from decompression to the SIMS analysis were carried out under cryogenic temperatures (Cryo-SIMS)[25] in order to minimize possible migration of hydrogen in the sample. In more detail[25], we decompressed the sample, fully opened the DAC, retrieved a rhenium gasket with the sample inside and placed the gasket in an in-house cryo-holder, all while immersed in liquid nitrogen. It was then transferred to a liquid nitrogen-cooled freezing stage in a SIMS instrument without being exposed to air. The temperature of the sample stage was monitored by a thermocouple and ensured to remain below –160 °C. Although it is difficult to precisely determine the temperature of the sample surface, the present measurements followed an analytical method[25] that retains hydrogen in metallic Fe, which requires at least –70 °C under vacuum[53]. In experiments made at >80 GPa, we almost always failed to recover a sample from the DAC in a liquid nitrogen pool, with an exception of run #9 performed at 117 GPa (however, the recovered sample in this run was so thin that we could not obtain the water content in melt that is milled by an ion beam more easily than surrounding Bdg and PPv). In seven out of thirteen runs, therefore, we quickly decompressed and recovered the sample in a glove box filled with $N_2$ gas and then transferred it to the cryogenic stage of the SIMS instrument, also without exposure to air.

Subsequently we examined the sample with Cryo-SIMS, an isotope microscope system, composed of a stigmatic SIMS instrument (CAMECA ims-1270e7) and a stacked CMOS-type active pixel sensor (SCAPS), which provides quantitative projection images of secondary ions emitted from the sample surface[54–56]. The sample surface was irradiated with a $^{16}O^-$ primary beam (23 keV, 200 nA), which was focused to 20–30 μm in diameter and rastered across a 150 μm × 150 μm region. The contrast aperture was set to be 150 μm in diameter. We offset the sample accelerating voltage by –100 V and employed an energy slit of 20 eV[57]. Pressure in the sample chamber was maintained at ~$4.5 \times 10^{-7}$ Pa during analyses. We performed pre-sputtering of the $^{16}O^-$ ion beam until a quenched melt pool was exposed to the surface, which was originally formed at the core of the sample (the hottest part) during laser heating (Extended Data Fig. 7).

Subsequently SCAPS images of secondary ions for $^{1}H^+$, $^{2}D^+$, $^{28}Si^+$, $^{56}Fe^+$, $^{40}Ca^+$, $^{27}Al^+$ and $^{24}Mg^+$ were collected with accumulation times of 100, 250, 15, 25, 25, 15 and 5 s, respectively (Extended Data Fig. 3). Since SCAPS exhibits a good linear relationship between the output voltage and the number of secondary ions, the abundance of each element and isotope can be quantified pixel by pixel from the intensity map[58]. The spatial resolution of quantification was 0.5 μm × 0.5 μm[38]. Three silicate glasses (including water-free one) with known $H_2O$ concentrations were used as standards (0.0–4.5 wt%)[47,59] to convert the intensity ratio, $^{1}H^+/^{28}Si^+$, into the mass ratio[38,47,48] (Extended Data Fig. 8a). The intensity ratios from secondary-ion images and the H/Si mass ratios of these standards have been reported to be linearly correlated[47,48] and was found to exhibit correlation coefficient $R^2 > 0.99$ in this study. $H_2O$ concentrations in Bdg, PPv and coexisting melt were then determined by multiplying the H/Si ratio by the Si content obtained by electron microprobe analyses normalized to total 100 wt% including water abundances (see below, Extended Data Table 1).

In addition, the $D_2O$ content was quantified by considering instrumental mass fractionation between hydrogen and deuterium. We synthesized an ($H_2O$, $D_2O$)-bearing rhyolitic glass by melting the JR-1 powder in an $Au_{75}Pd_{25}$ capsule at 1.0 GPa and 1,573 K in a piston-cylinder apparatus. In order to homogenize $H_2O$ and $D_2O$ concentrations in the glass, the sample was once melted for 2 hr under such *P*-*T* conditions, recovered, ground in a mortar, and then melted for another 1 hr at 1 GPa. The D/H ratio of this synthesized glass was quantified to be 0.00331(3) by an isotope ratio mass spectrometer (IRMS) (DELTA V Advantage, Thermo Fisher Scientific) coupled with a thermal conversion elemental analyzer (TC/EA) at Geo-Science Laboratory Co. Ltd., Nagoya, Japan. The instrumental mass fractionation, IMF (‰) = $1000[(D/H)_{SIMS}/(D/H)_{true} - 1]$, for D/H was examined by measuring $(D/H)_{SIMS}$ of this glass standard while varying the sample offset voltage (ion energy) with an energy slit of 20 eV fixed (Extended Data Fig. 8b). The results demonstrate that the D/H IMF value is sensitive to the sample offset voltage and linearly increases for a larger offset (corresponding to a larger ion energy) as reported by Hauri *et al.*[27]. It indicates that the sample offset voltage of –100 V in our analyses helps not only eliminate hydrogen not intrinsic to the sample[57] but also diminish the magnitude of D/H IMF. We also found that the D/H IMF is –266(8)‰ under the present analytical conditions, close to the value reported in Hauri *et al.*[27]. The $D_2O$ content was quantified based on a combination of the calibration curve for H/Si (Extended Data Fig. 8a) and this D/H IMF value.

**Matrix effect in hydrogen isotope analysis by SIMS**

The present experiments demonstrate large hydrogen isotopic fractionation between Bdg and melt with isotopic fractionation factor, $\alpha = (D/H)_{Bdg}/(D/H)_{melt} = 1.15–1.30$ (Fig. 3, Extended Data Table 1). It is significant even when matrix effects in hydrogen isotope analysis are taken into account[27,28]. Since $\alpha = (D_{Bdg}/D_{melt})/(H_{Bdg}/H_{melt})$ by formula conversion, differences in the relative sensitivity coefficient between hydrogen and deuterium do not affect the determination of α (in other words, precise determination of $D_2O$ concentration is not necessary to obtain α). Furthermore, we collected two-dimensional secondary ion images simultaneously for Bdg and melt using the SCAPS system, which ensures that secondary ions from Bdg and melt possessed an identical energy although the D/H IMF value changes with varying the ion energy as demonstrated in Extended Data Fig. 8b.

**Electron microprobe analyses**

After SIMS measurements, we performed textural characterizations and major element analyses of a sample with a field emission-type scanning electron microscope (FE-SEM, JSM-7000F, JEOL), which is equipped with an energy-dispersive X-ray spectrometer (EDS) using a silicon drift detector (X-Max 150, Oxford). The EDS analyses with an acceleration voltage of 15 kV and a beam current of 1 nA provided X-ray elemental maps (Fig. 1 and Extended Datra Fig. 3) and the major element compositions of quenched partial melt and surrounding Bdg and/or PPv (Extended Data Table 1). We examined the sample surface exposed by ion beam sputtering during the SIMS measurements. For samples whose surface was not smooth, we obtained its cross section with a focused Ga ion beam (FIB; Thermo Scientific, Helios 5 UC Dualbeam) and determined their major element compositions with an FE-type electron probe microanalyzer (FE-EPMA; JEOL, JXA-8530F) employing 12 kV acceleration voltage, 10 nA beam current, analysing crystals of PETJ (Si, Ca, K), TAP (Al, Mg), LIF (Fe) and TAPH (Na) and standards of $SiO_2$, $Al_2O_3$, Fe, MgO, $CaSiO_3$, $NaAlSi_2O_6$ and $KTiOPO_4$.

**Deuterium loss from quenched melt**

We employ the present analyses of melt $D_2O$ concentrations only for samples that were decompressed under cryogenic temperatures in liquid nitrogen (Extended Data Table 1).

It is because the measured melt $D_2O$ abundances (Extended Data Fig. 4d) and $D_{D2O}$(Bdg/melt) values (Extended Data Figs. 4e, f) are clearly distinguished between samples decompressed/recovered under cryogenic temperatures in liquid nitrogen ($D_{D2O}$ = 0.03–0.11, 0.5–2.9 wt% $D_2O$ in melt) and room temperature in nitrogen gas ($D_{D2O}$ = 0.3–3.9, 0.0–0.4 wt% $D_2O$ in melt), indicating that most of the deuterium in quenched silicate melt was lost when decompressed at ambient temperature. We also note that while the sample was decompressed in liquid nitrogen in run #2, the sample temperature was increased in a SIMS instrument because the supply of liquid nitrogen was suspended. As a consequence, the $D_{D2O}$(Bdg/melt) value obtained in this experiment was remarkably higher than $D_{H2O}$(Bdg/melt) unlike other runs recovering a sample at cryogenic temperatures (Extended Data Fig. 4f), which is likely attributed to a partial loss of deuterium from melt.

Except $D_2O$ in melt, water in Bdg, PPv and melt should have been preserved until Cryo-SIMS measurements even when sample was decompressed at 300 K. The measured $D_{H2O}$(Bdg/melt) values are consistent among all runs in this study (Fig. 2b) except run #4 (due to water saturation in Bdg, see the main text), regardless of the temperatures during decompression. When using the data only from runs with cryogenic sample recovery, we obtain pressure and temperature dependences for $D_{H2O}$(Bdg/melt) similar to those based on all the dataset (Extended Data Fig. 9). In addition, $D_2O$ concentrations in Bdg (or PPv) are comparable between samples recovered under cryogenic and room temperatures (Extended Data Fig. 4c).

**Pressure of the phase transition between Bdg and PPv**

The pressure and temperature range for stability of PPv in a molten mantle remains unclear[32,50], in particular when it is enriched in FeO and $H_2O$ relevant to an evolved BMO. For a pyrolitic representative mantle material, recent melting experiments[50] demonstrated the formation of PPv above ~120 GPa along the solidus curve (Extended Data Fig. 1a) (note that their sample could have included ~1,500 ppm $H_2O$ due to adsorption of water from air, see Fig. S5 in Nomura *et al.*[60] who used a similar starting material). On the other hand, the dry experiments conducted by Fiquet *et al.*[32] observed that PPv underwent a phase transition to Bdg with increasing temperature to 3,500 K at 138 GPa, suggesting that PPv just barely does not appear above solidus temperatures within the pressure range of a dry Earth mantle (<136 GPa). Even at modest deep lower-mantle temperatures of ~2,400 K, while the Bdg-PPv transition has been repeatedly reported to occur in a pyrolitic lowermost mantle[61–63], Grocholski *et al.*[64] did not find PPv below ~140 GPa.

Crystallization in a BMO leads to enrichment in both FeO and $H_2O$ in the residual melts, both of which stabilize PPv with respect to Bdg. The present water-bearing experiments observed the coexistence of PPv, Bdg and melt at 117 GPa and 4,240 K (run #9), showing a slightly wider stability field of PPv than in Kuwayama *et al.*'s[50] experiments (Extended Data Fig. 1a). It has been also demonstrated[65] that PPv appears in the $Fe_2SiO_4$ sample above 95 GPa at 2,400 K, consistent with earlier experiments[66,67] and theoretical predictions[67,68]. Recent calculations[12] found that the Bdg-PPv transition pressure for $MgSiO_3$ is reduced by 10 GPa when incorporating 0.56 wt% $H_2O$ by $Si^{4+} = Al^{3+} + H^+$ substitution. In the present modelling of BMO solidification, we therefore considered the formation of hydrous PPv from an (FeO, $H_2O$)-enriched BMO above ~114 GPa (Extended Data Fig. 1b) after 91% solidification (Fig. 4c).

## Modelling the evolution of a basal magma ocean

We modelled the solidification of a BMO following Ozawa *et al.*[44] to examine the evolution of $H_2O$ and $D_2O$ concentrations and the resulting D/H ratio in residual BMO melts and crystallizing Bdg and PPv. Our model considers that the solidification of a magma ocean proceeds via batch crystallization until a rheological transition[69] occurs at the residual melt fraction $F$, which also controls the depth of the density crossover between melt and Bdg[70] and therefore defines the mass fraction $M_{\mathrm{bdc}}$ of the entire mantle involved in the chemical evolution of the BMO including the initial batch crystallization as[44]:

$$M_{\mathrm{bdc}} = 0.981 \times F^2 - 2.28 \times F + 1.38 \tag{4}$$

Here we assume $F = 0.5$ by following ref. 70, which leads to melt-Bdg density crossover at 50 GPa[70] and the initial BMO extending from 85 to 136 GPa by collecting melts that segregate downward after 50% batch crystallization only of Bdg at 50–136 GPa. The BMO then evolves by fractional crystallization of Bdg:ferropericlase(Fp)=6:4 by weight at its top. Fp is assumed to be completely anhydrous. The *P-T* conditions of batch crystallization and the *P-T* path for fractional crystallization are shown in Extended Data Fig. 1b, in which the temperature of the present-day ULVZs, equivalent to the CMB temperature, is assumed to be 3,800 K. We set the Bdg-PPv transition boundary in the BMO based primarily on the result of run #9 demonstrating the three-phase coexistence of Bdg, PPv and melt (Extended Data Fig. 1a), across which the crystallizing phase switches from Bdg to PPv. We adopt the PREM model[71] for the depth-pressure relation. The presence of an atmosphere is not considered.

After 50% batch crystallization, the total mass of hydrogen in the BMO melt, $\mathrm{H_{f0}}$, is given by:

$$\mathrm{H_{f0}} = \mathrm{H_{BSE}} \left\{ \int_{M_{\mathrm{BPb}}^{\mathrm{batch}}}^{M_{\mathrm{bdc}}} \frac{F}{D_{\mathrm{H}}^{\mathrm{Bdg/melt}}(1-F)+F} dm + \int_{0}^{M_{\mathrm{BPb}}^{\mathrm{batch}}} \frac{F}{D_{\mathrm{H}}^{\mathrm{PPv/melt}}(1-F)+F} dm \right\} \tag{5}$$

in which $\mathrm{H_{BSE}}$ represents the total mass of hydrogen included in the initial fully molten mantle after core metal segregations, $M_{\mathrm{BPb}}^{\mathrm{batch}}$ is the mass fraction below the Bdg-PPv boundary to the entire mantle, and $D_{\mathrm{H}}^{\mathrm{Bdg/melt}}$ and $D_{\mathrm{H}}^{\mathrm{PPv/melt}}$ are the Bdg/melt and PPv/melt partition coefficients of hydrogen ($H_2O$), respectively, under *P-T* conditions at a given depth, the mass fraction below which is given by $m$.

In the subsequent fractional crystallization, the total mass of hydrogen in the BMO residual melt is expressed as a function of BMO weight fraction relative to the entire mantle, $M_m$ as:

$$\mathrm{H}_m(M_m) = \begin{cases} \mathrm{H_{f0}} \exp\left( -\int_{M_m}^{0.5M_{\mathrm{bdc}}} \frac{0.6D_{\mathrm{H}}^{\mathrm{Bdm/melt}}}{m} dm \right) \ (M_m \geq M_{\mathrm{BPb}}^{\mathrm{frac}}) \\ \mathrm{H_{f0}} \exp\left( -\int_{M_m}^{M_{\mathrm{BPb}}^{\mathrm{frac}}} \frac{0.6D_{\mathrm{H}}^{\mathrm{PPv/melt}}}{m} dm - \int_{M_{\mathrm{BPb}}^{\mathrm{frac}}}^{0.5M_{\mathrm{bdc}}} \frac{0.6D_{\mathrm{H}}^{\mathrm{Bdm/melt}}}{m} dm \right) \ (M_m < M_{\mathrm{BPb}}^{\mathrm{frac}}) \end{cases} \tag{6}$$

Changes in the proportion of hydrogen in the BMO residual melt with respect to $\mathrm{H_{BSE}}$ is illustrated in Fig. 4a, showing 0.352 when $M_m = 0.02$ corresponding to the size of the observed LLSVPs. Considering that 35.2% of $H_2O$ is sequestrated in the BMO that could have further evolved into the less commonly sampled LLSVPs and ULVZs, the remaining 64.8% yields the observed 710 ppm $H_2O$ in the complementary reservoir[46], suggesting that the original $H_2O$ content in a fully molten mantle before the onset of crystallization is 1,075 ppm. This yields 1.89 wt% $H_2O$ in the BMO at $M_m = 0.02$ (Fig. 4b). Hydrogen concentrations in solid Bdg and PPv $C_s^{\mathrm{H}}$, crystallizing from the BMO are also calculated as a function of $M_m$ as:

$$C_s^{\mathrm{H}}(M_m) = \begin{cases} D_{\mathrm{H}}^{\frac{\mathrm{Bdg}}{\mathrm{melt}}} \frac{\mathrm{H}_m(M_m)}{M_m \widetilde{M}_{\mathrm{BSE}}} & (M_m \geq M_{\mathrm{BPb}}^{\mathrm{frac}}) \\ D_{\mathrm{H}}^{\frac{\mathrm{PPv}}{\mathrm{melt}}} \frac{\mathrm{H}_m(M_m)}{M_m \widetilde{M}_{\mathrm{BSE}}} & (M_m < M_{\mathrm{BPb}}^{\mathrm{frac}}) \end{cases} \quad (7)$$

where $\widetilde{M}_{\mathrm{BSE}}$ is the entire mantle mass (Fig. 4c). For the evolution of the D/H ratio in the BMO residual melt and crystallizing Bdg/PPv (Fig. 4d), similar calculations are applied to obtain $\mathrm{D}_m(M_m)$ and $C_s^{\mathrm{D}}(M_m)$ for deuterium using its crystals/melt partition coefficient $D_{\mathrm{D}}^{\mathrm{Bdg(PPv)/melt}} = D_{\mathrm{H}}^{\mathrm{Bdg(PPv)/melt}} \times \alpha$, in which α is hydrogen isotopic fractionation factor (Eq. 3).

To estimate uncertainties in $\mathrm{H}_m(M_m)$, $\mathrm{D}_m(M_m)$, $C_s^{\mathrm{H}}(M_m)$ and $C_s^{\mathrm{D}}(M_m)$ considering errors in parameters for $D_{H_2O}$(Bdg/melt) (Eq. 1) and α (Eq. 3), we employed a Monte Carlo approach with 100,000 random samples from a multivariate normal distribution, which is defined by the regression-derived parameters and their covariance matrix (variances: constant term = 0.190, $1/T$ term = $1.94 \times 10^6$ and $P/T$ term = 76.2; covariances: constant–$1/T$ = –591, constant–$P/T$ = –2.97 and $1/T$–$P/T$ = $7.65 \times 10^3$). The nominal value of each plot in Fig. 4 indicates the average of the Monte Carlo samples, and their lower and upper bounds represent the 16th and 84th percentiles, respectively.

**Water abundance and the D/H ratio of an initial magma ocean**

According to Hirschmann *et al.*[46], the accessible (and continuously sampled) near-surface and mantle reservoirs (complement of the BMO water reservoir) contain $2{,}850 \times 10^{21}$ g $H_2O$ (710±90 ppm by weight), consisting mainly of convecting mantle ($1{,}200 \times 10^{21}$ g $H_2O$), lithospheric mantle ($50 \times 10^{21}$ g $H_2O$), crust (sedimentary rocks, $200 \times 10^{21}$ g $H_2O$)[72] and oceans ($1{,}400 \times 10^{21}$ g $H_2O$). We calculated the hydrogen isotopic composition of the entire complementary reservoir to be δD = –38‰, considering the contributions of convecting and lithospheric mantle (δD = –75‰)[73], sedimentary rocks (δD = –75‰)[72] and oceans (δD = 0‰).

Along with these values for the complementary reservoir, considering that 35.2% of $H_2O$ in the initial magma ocean is now present in the LLSVPs and ULVZs (see above, Fig. 4a) with δD = –124‰ on average (Fig. 4d), the fully molten mantle (initial magma ocean before crystallization but after core formation) may have contained 1,075 ppm $H_2O$ with initial δD = –67.6‰.

**Data availability**

The authors declare that the data supporting the findings of this study are available within the paper.

**Code availability**

The codes to reproduce this paper are available from the corresponding author upon reasonable request.

49. Takahashi, E. Melting of a dry peridotite KLB-1 up to 14 GPa: Implications on the origin of peridotitic upper mantle. *J. Geophys. Res. Solid Earth* **91**, 9367–9382 (1986).
50. Kuwayama, Y. *et al.* Post-perovskite phase transition in the pyrolitic lowermost mantle: Implications for ubiquitous occurrence of post-perovskite above CMB. *Geophys. Res. Lett.* **49**, e2021GL096219 (2022).
51. Tateno, S. *et al.* Melting phase relations and element partitioning in MORB to lowermost mantle conditions. *J. Geophys. Res. Solid Earth* **123**, 5515–5531 (2018).
52. Mori, Y. *et al.* Melting experiments on Fe–$Fe_3S$ system to 254 GPa. *Earth Planet. Sci. Lett.* **464**, 135–141 (2017).
53. Antonov, V. E. *et al.* Solubility of deuterium and hydrogen in fcc iron at high pressures and temperatures. *Phys. Rev. Materials* **3**, 113604 (2019).
54. Greenwood, J. P. *et al.* Hydrogen isotope ratios in lunar rocks indicate delivery of cometary water to the Moon. *Nat. Geosci.* **4**, 79–82 (2011).
55. Sakamoto, N. *et al.* Remnants of the early solar system water enriched in heavy oxygen isotopes. *Science* **317**, 231–233 (2007).
56. Yurimoto, H., Nagashima, K. & Kunihiro, T. High precision isotope micro-imaging of materials. *Appl. Surf. Sci.* **203-204**, 793–797 (2003).
57. Yurimoto, H., Kurosawa, M. & Sueno, S. Hydrogen analysis in quartz crystals and quartz glasses by secondary ion mass spectrometry. *Geochim. Cosmochim. Acta* **53**, 751–755 (1989).
58. Yamamoto, K., Sakamoto, N. & Yurimoto, H. Analysis of the noise properties of a solid-state SCAPS ion imager and development of software noise reduction. *Surf. Interface Anal.* **42**, 1603–1605 (2010).
59. Yoshimura, S. Diffusive fractionation of $H_2O$ and $CO_2$ during magma degassing. *Chem. Geol.* **411**, 172–181 (2015).
60. Nomura, R. *et al.* Low core-mantle boundary temperature inferred from the solidus of pyrolite. *Science* **343**, 522–525 (2014).
61. Murakami, M., Hirose, K., Sata, N. & Ohishi, Y. Post-perovskite phase transition and mineral chemistry in the pyrolitic lowermost mantle. *Geophys. Res. Lett.* **32**, 2004GL021956 (2005).
62. Ohta, K. *et al.* Electrical conductivities of pyrolitic mantle and MORB materials up to the lowermost mantle conditions. *Earth Planet. Sci. Lett.* **289**, 497–502 (2010).
63. Ono, S. & Oganov, A. In situ observations of phase transition between perovskite and $CaIrO_3$-type phase in $MgSiO_3$ and pyrolitic mantle composition. *Earth Planet. Sci. Lett.* **236**, 914–932 (2005).
64. Grocholski, B., Catalli, K., Shim, S.-H. & Prakapenka, V. Mineralogical effects on the detectability of the postperovskite boundary. *Proc. Natl. Acad. Sci. U.S.A.* **109**, 2275–2279 (2012).
65. Yang, Z., Song, Z., Wu, Z., Mao, H. & Zhang, L. Iron silicate perovskite and postperovskite in the deep lower mantle. *Proc. Natl. Acad. Sci. U.S.A.* **121**, e2401281121 (2024).
66. Dorfman, S. M., Meng, Y., Prakapenka, V. B. & Duffy, T. S. Effects of Fe-enrichment on the equation of state and stability of $(Mg,Fe)SiO_3$ perovskite. *Earth Planet. Sci. Lett.* **361**, 249–257 (2013).
67. Mao, W. L. *et al.* Iron-rich silicates in the Earth's D″ layer. *Proc. Natl. Acad. Sci.*

*U.S.A.* **102**, 9751–9753 (2005).
68. Stackhouse, S., Brodholt, J. P. & Price, G. D. Elastic anisotropy of $FeSiO_3$ end-members of the perovskite and post-perovskite phases. *Geophys. Res. Lett.* **33**, 2005GL023887 (2006).
69. Costa, A., Caricchi, L. & Bagdassarov, N. A model for the rheology of particle-bearing suspensions and partially molten rocks. *Geochem. Geophys. Geosyst.* **10**, 2008GC002138 (2009).
70. Caracas, R., Hirose, K., Nomura, R. & Ballmer, M. D. Melt–crystal density crossover in a deep magma ocean. *Earth Planet. Sci. Lett.* **516**, 202–211 (2019).
71. Dziewonsk, A. M. & Anderson, D. L. Preliminary reference Earth model. *Phys. Earth Planet. Inter.* **25**, 297–356 (1981).
72. Lécuyer, C., Gillet, P. & Robert, F. The hydrogen isotope composition of seawater and the global water cycle. *Chem. Geol.* **145**, 249–261 (1998).
73. Loewen, M. W., Graham, D. W., Bindeman, I. N., Lupton, J. E. & Garcia, M. O. Hydrogen isotopes in high $^3He/^4He$ submarine basalts: Primordial vs. recycled water and the veil of mantle enrichment. *Earth Planet. Sci. Lett.* **508**, 62–73 (2019).
74. Kim, T. *et al.* Low melting temperature of anhydrous mantle materials at the core-mantle boundary. *Geophys. Res. Lett.* **47**, e2020GL089345 (2020).

**Acknowledgements** Synchrotron XRD measurements were performed at the beamline BL10XU, SPring-8 (proposals nos. 2024B1200, 2025A1133, 2025B1212 and 2025B1699). We thank K. Yonemitsu for EPMA analyses. Discussion with H. Yurimoto and J. Tsuchiya was helpful. This work was supported by JSPS Kakenhi 21H04968 and 26H02080 to K.H. and JST FOREST JPMJFR2173 to N.S.

**Author Contributions** K.H. designed and led the project. This study is based on DAC experiments by M.S. and Y.T., SIMS analyses by N.S., Shu.M. and Shi.M., and computational modelling by K.O. K.H., M.S. and J.W.H. discussed implications of the present results and wrote the manuscript on which all authors commented.

**Competing interests** The authors declare no competing interests.

**Correspondence and requests for materials** should be addressed to Megumi Semba or Kei Hirose.

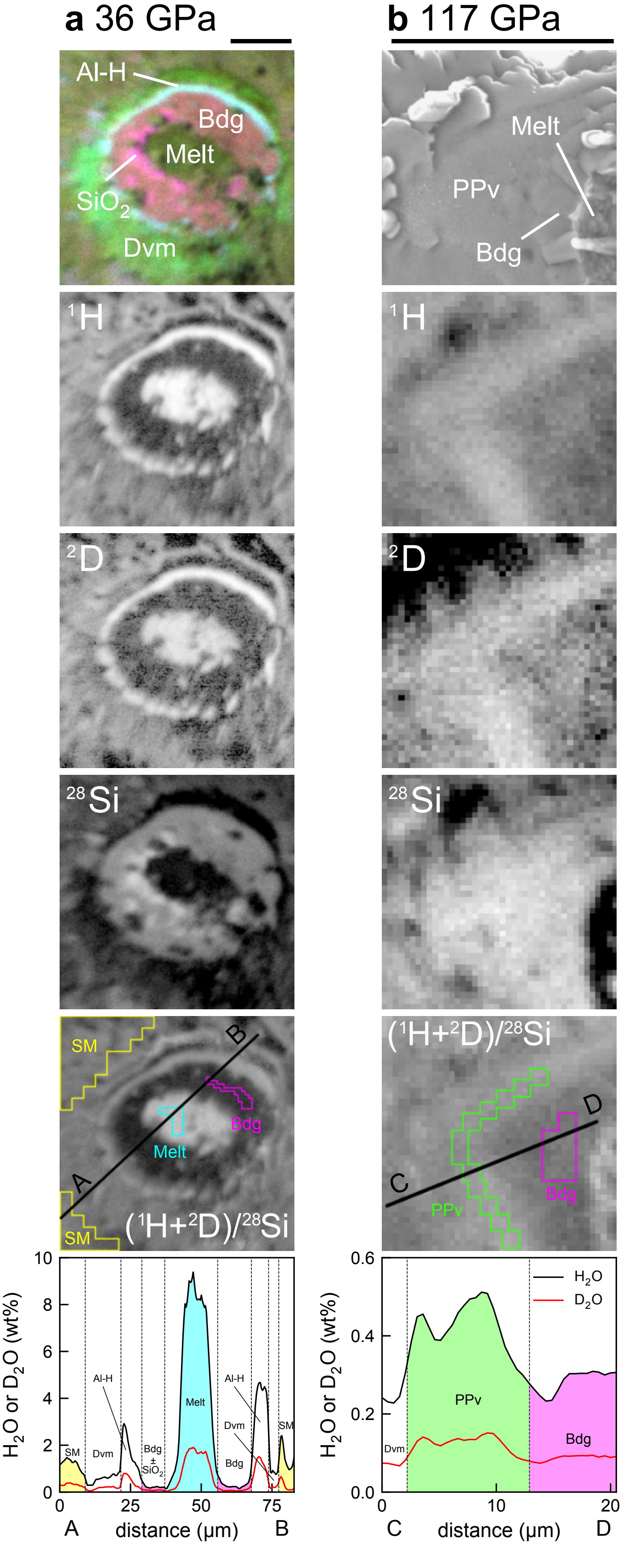
a 36 GPa
b 117 GPa
Al-H
Bdg
Melt
SiO2
Dvm
PPv
$^1$H
$^2$D
$^{28}$Si
SM
B
A
Melt
Bdg
($^1$H+$^2$D)/$^{28}$Si
C
D
PPv
H2O or D2O (wt%)
distance (μm)
H2O
D2O
Dvm
Al-H

**Fig. 1 | Cryo-SIMS analyses of $H_2O$ and $D_2O$ in Bdg/PPv and coexisting melt.** Samples were synthesized at (**a**) 36 GPa and 3,340 K (run #1) and (**b**) 117 GPa and 4,240 K (run #9) and decompressed/recovered/SIMS-analysed under cryogenic temperatures. Uppermost panels: (**a**) combined secondary ion images of Mg (yellow), Si (magenta), Ca (green) and Al (cyan) exhibits quenched melt at the centre, surrounded by Bdg, Al-H phase and davemaoite (Dvm) layers. A small amount of the $SiO_2$ phase was present within the Bdg layer. The outside of the layered region was not involved in melting, and thus its composition was unchanged from the starting material (SM). (**b**) backscattered electron image showing the Bdg and PPv single-phase layers surrounding quenched melt. Bdg crystals exhibit a euhedral shape, whereas PPv became amorphous during decompression to ambient pressure. Middle panels: the secondary ion images of $^{1}H^{+}$, $^{2}D^{+}$, $^{28}Si^{+}$ and $(^{1}H^{+}+^{2}D^{+})/^{28}Si^{+}$. The $H_2O$ and $D_2O$ contents for each phase were obtained for the region of interest (ROI) shown in the $(^{1}H^{+}+^{2}D^{+})/^{28}Si^{+}$ image. We avoided the area within two or more pixels from phase boundaries. Lowermost panels: profiles of $H_2O$ and $D_2O$ concentrations along the lines in $(^{1}H^{+}+^{2}D^{+})/^{28}Si^{+}$ maps. All images are shown using a logarithmic scale. Scale bars, 20 μm.

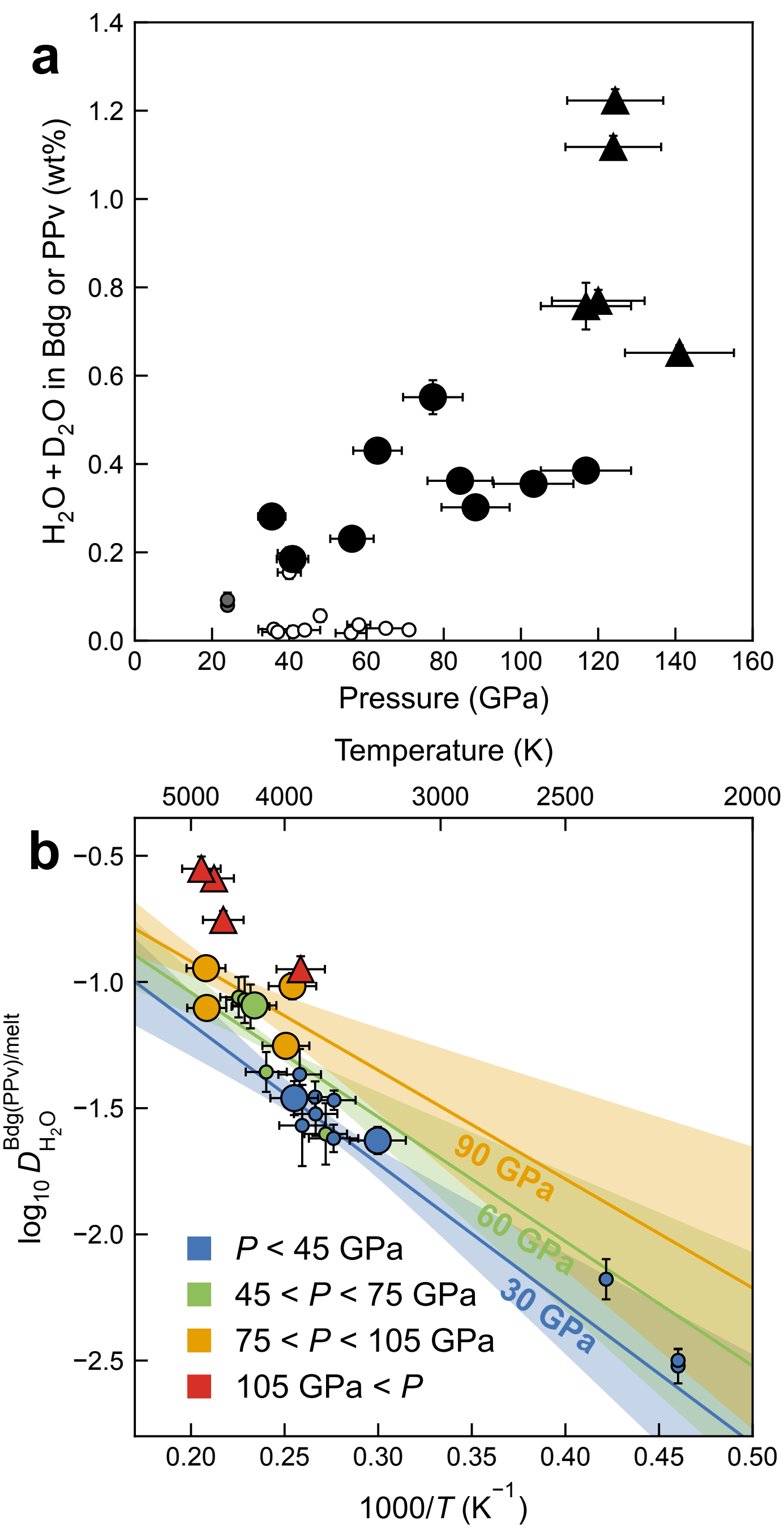

**Fig. 2 | Water concentrations in Bdg and PPv and its partitioning with melt.** Circles, Bdg; triangles, PPv. Large and small symbols are from this study and earlier experiments, respectively. **a,** Variations in the SIMS analyses of water ($H_2O + D_2O$ combined, wt%). Grey, ref. 19; open, ref. 10. **b,** Logarithm of Bdg(PPv)/melt partition coefficients of $H_2O$ (weight basis) as a function of reciprocal temperature. A coloured line represents the temperature dependence at respective pressure (blue, 30 GPa; green, 60 GPa; orange, 90 GPa), based on data from this study and ref. 10 (Eq. 1). The shaded areas show 1σ uncertainty of the fits. The fitting results are consistent with data previously reported by using a large-volume press at 24 GPa[19].

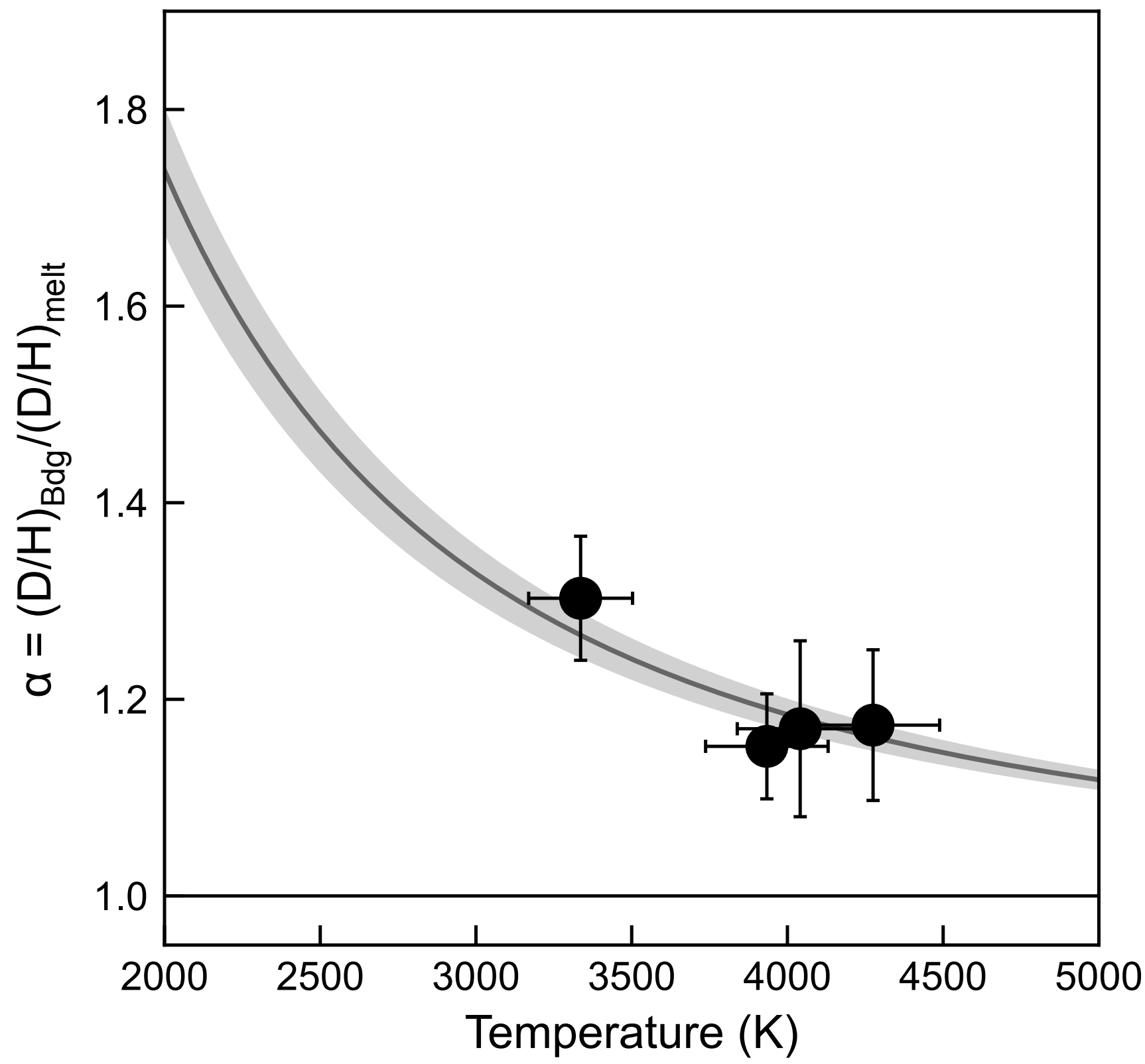

**Fig. 3 | Hydrogen isotopic fractionation factor between Bdg and melt, $\alpha = (D/H)_{Bdg}/(D/H)_{melt}$, observed at 36–77 GPa.** The curve shows the conventional equation for temperature-dependence of the isotopic fractionation factor (Eq. 3). The shaded band indicates 1σ uncertainty. The uncertainty at low temperatures may be larger than illustrated here since Eq. 3 approximates the temperature effect at relatively high temperatures[29,30].

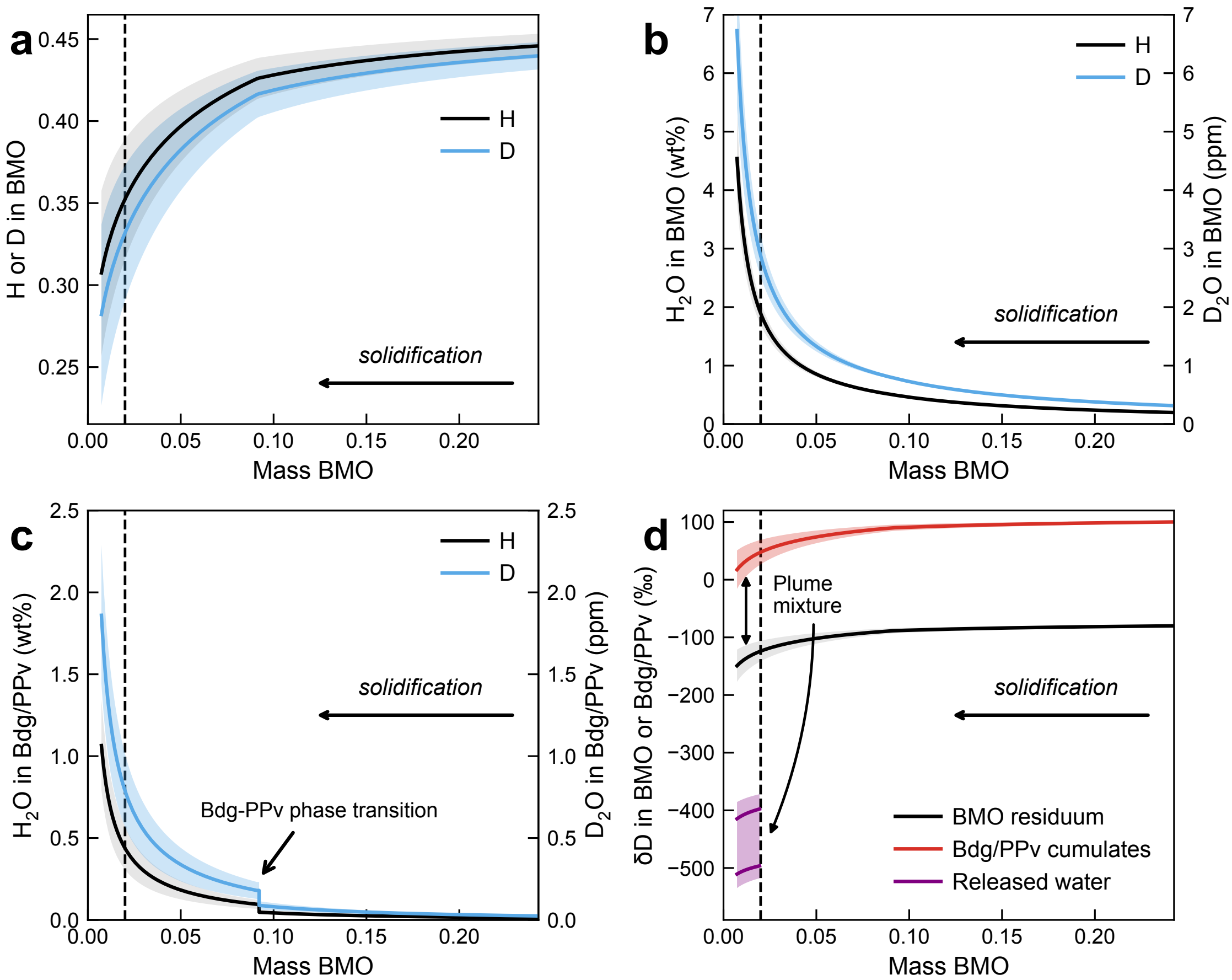

**Fig. 4 | Evolution of water abundances and hydrogen isotopic compositions of BMO and crystallizing Bdg and PPv.** Each graph is given as a function of the mass fraction of BMO residual melt relative to the entire mantle. Temporal evolution (solidification) is right-to-left in all panels. A vertical broken line indicates the time when the amount of BMO melt reduces to 2% of the mantle (corresponding to the size of present day LLSVPs). **a,** Proportions of $H_2O$ (black) and $D_2O$ (blue) present in the BMO to those in the bulk silicate Earth including oceans. **b,c,** $H_2O$ (wt%) and $D_2O$ (ppm) concentrations in the BMO (**b**) and in crystallizing Bdg and PPv (**c**) that appear before and after the BMO mass fraction reaches 9%, respectively. **d,** Hydrogen isotopic compositions of BMO residuum (black) and crystallizing Bdg and PPv (red). The purple range represents D-poor incipient water that would be released from a mixture of hydrous BMO residuum and hydrous Bdg/PPv entrained in an upwelling mantle plume at shallow lower mantle depths, which may explain reported negative δD values in plume-derived melts[7,8]. It is calculated from the δD of hydrous PPv (red) with α = 1.7 at 2,000 K (Fig. 3) considering Bdg dehydration at the middle/shallow lower mantle and also from that of the BMO residuum (black) when it solidifies and releases water to surrounding PPv/Bdg in upwellings, which dehydrates at shallower depths accompanying isotopic fractionation. See the main text and Methods for details.

**Extended Data Table 1 | Experimental results**

| Run number | 1* | 2* | 3* | 4*† | 5* | 6‡ | 7 | 8‡ | 9* | | 10 | 11 | 12‡ | 13 |
|---|---|---|---|---|---|---|---|---|---|---|---|---|---|---|
| Pressure (GPa) | 36(4) | 41(4) | 56(6) | 63(6) | 77(8) | 84(8) | 88(9) | 103(10) | 117(12) | | 120(12) | 124(12) | 124(12) | 141(14) |
| Temperature (K) | 3,340(170) | 3,920(200) | 4,280(210) | 4,040(200) | 3,930(200) | 3,990(200) | 4,800(240) | 4,810(240) | 4,240(210) | | 4,710(240) | 3,870(190) | 4,860(240) | 4,600(230) |
| | | | | | | | | | | | | | | |
| Phase | Bdg | Bdg | Bdg | Bdg | Bdg | Bdg | Bdg | Bdg | Bdg | PPv | PPv | PPv | PPv | PPv |
| $SiO_2$ | 56.43(77) | 56.52(158) | 57.32(106) | 58.00(50) | 54.68(151) | 57.56(37) | 57.45(48) | 54.48(46) | 56.40(89) | 53.13(168) | 55.11(93) | 55.36(105) | 53.26(20) | 52.64(91) |
| $Al_2O_3$ | 2.28(13) | 2.91(22) | 2.76(22) | 2.97(19) | 2.80(20) | 2.35(5) | 2.87(8) | 1.86(21) | 3.10(15) | 3.09(22) | 2.61(18) | 3.47(15) | 4.69(10) | 3.57(18) |
| FeO | 1.08(11) | 1.64(11) | 0.84(15) | 1.63(12) | 0.95(35) | 0.53(14) | 0.90(5) | 0.88(7) | 0.48(21) | 2.10(55) | 0.70(14) | 0.40(6) | 0.06(1) | 0.21(5) |
| MgO | 39.98(104) | 38.56(104) | 38.91(106) | 37.18(66) | 40.84(68) | 39.12(24) | 38.15(24) | 42.19(33) | 39.57(71) | 38.97(80) | 41.00(91) | 39.31(94) | 41.67(41) | 42.42(79) |
| CaO | 0.23(4) | 0.36(5) | 0.18(8) | 0.22(2) | 0.72(22) | 0.39(4) | 0.63(5) | 0.56(1) | 0.45(4) | 2.70(64) | 0.59(6) | 1.46(30) | 0.20(1) | 1.16(23) |
| $Na_2O$ | | | | | | 0.05(1) | | 0.03(1) | | | | | 0.12(1) | |
| $H_2O$ | 0.217(17) | 0.139(12) | 0.185(7) | 0.339(15) | 0.469(38) | 0.314(8) | 0.206(15) | 0.274(22) | 0.285(22) | 0.569(49) | 0.634(24) | 0.861(24) | 0.945(23) | 0.477(16) |
| $D_2O$ | 0.065(5) | 0.046(4) | 0.046(2) | 0.091(6) | 0.083(7) | 0.048(2) | 0.095(7) | 0.081(6) | 0.100(8) | 0.189(19) | 0.136(5) | 0.257(8) | 0.278(11) | 0.174(7) |
| $H_2O+D_2O$ | 0.281(17) | 0.185(13) | 0.231(7) | 0.430(16) | 0.551(38) | 0.362(8) | 0.302(17) | 0.355(23) | 0.385(23) | 0.757(53) | 0.770(25) | 1.118(25) | 1.223(26) | 0.652(17) |
| Total | 99.19 | 96.58 | 90.2 | 95.18 | 96.5 | 98.63 | 98.44 | 97.81 | 97.23 | 95.08 | 99.61 | 90.79 | 98.28 | 90.13 |
| | | | | | | | | | | | | | | |
| Phase | Melt | Melt | Melt | Melt | Melt | Melt | Melt | Melt | Melt | | Melt | Melt | Melt | Melt |
| $SiO_2$ | 30.66(197) | 25.60(267) | 27.94(213) | 33.83(167) | 23.32(76) | 40.89(177) | 17.21(52) | 23.82(35) | 43.68(378) | | 22.08(119) | 28.31(300) | 11.39(94) | 30.79(196) |
| $Al_2O_3$ | 3.27(18) | 2.84(29) | 3.00(21) | 4.85(24) | 5.00(43) | 6.74(36) | 3.45(10) | 3.56(9) | 3.71(12) | | 4.34(32) | 4.56(15) | 3.19(9) | 4.45(14) |
| FeO | 5.18(68) | 10.71(189) | 25.27(390) | 9.63(65) | 4.88(65) | 3.98(29) | 35.48(269) | 14.80(44) | 5.53(237) | | 0.79(6) | 1.66(28) | 0.97(7) | 1.20(10) |
| MgO | 57.19(178) | 53.01(403) | 39.77(158) | 45.38(226) | 58.11(68) | 37.71(139) | 40.71(99) | 50.35(85) | 44.86(106) | | 62.66(383) | 56.85(175) | 67.18(95) | 55.36(276) |
| CaO | 3.70(43) | 7.84(64) | 4.02(33) | 6.31(116) | 8.69(18) | 10.14(28) | 3.16(10) | 7.30(8) | 2.22(59) | | 10.13(27) | 8.62(67) | 16.48(34) | 8.19(54) |
| $Na_2O$ | | | | | | 0.54(2) | | 0.17(5) | | | | | 0.78(2) | |
| $H_2O$ | 9.200(896) | 4.016(514) | 2.286(196) | 12.740(742) | 4.867(431) | 5.622(457) | 2.617(204) | 2.414(184) | | | 2.464(163) | 7.660(883) | 3.364(365) | 2.710(204) |
| $D_2O$ | 2.106(206) | | 0.487(43) | 2.917(177) | 0.745(67) | | | | | | | | | |
| $H_2O+D_2O$ | 11.305(919) | 4.710(521) | 2.773(201) | 15.657(762) | 5.612(437) | 5.762(457) | 2.643(204) | 2.435(184) | | | 2.534(164) | 7.942(885) | 3.573(367) | 3.121(208) |
| Total | 104.61 | 96.11 | 83.44 | 91.39 | 102.52 | 77.27 | 80.09 | 83.75 | 90.83 | | 94.38 | 88.23 | 89.44 | 87.57 |
| | | | | | | | | | | | | | | |
| $D$ ($H_2O$) | 0.024(3) | 0.035(5) | 0.081(8) | 0.027(2) | 0.096(12) | 0.056(5) | 0.079(8) | 0.113(12) | | | 0.257(20) | 0.112(13) | 0.281(31) | 0.176(14) |
| $\alpha$ | 1.30(6) | | 1.17(8) | 1.17(9) | 1.15(5) | | | | | | | | | |

*Sample was decompressed in liquid nitrogen

†Bdg was saturated with $H_2O$

‡EPMA analysis (except for water)

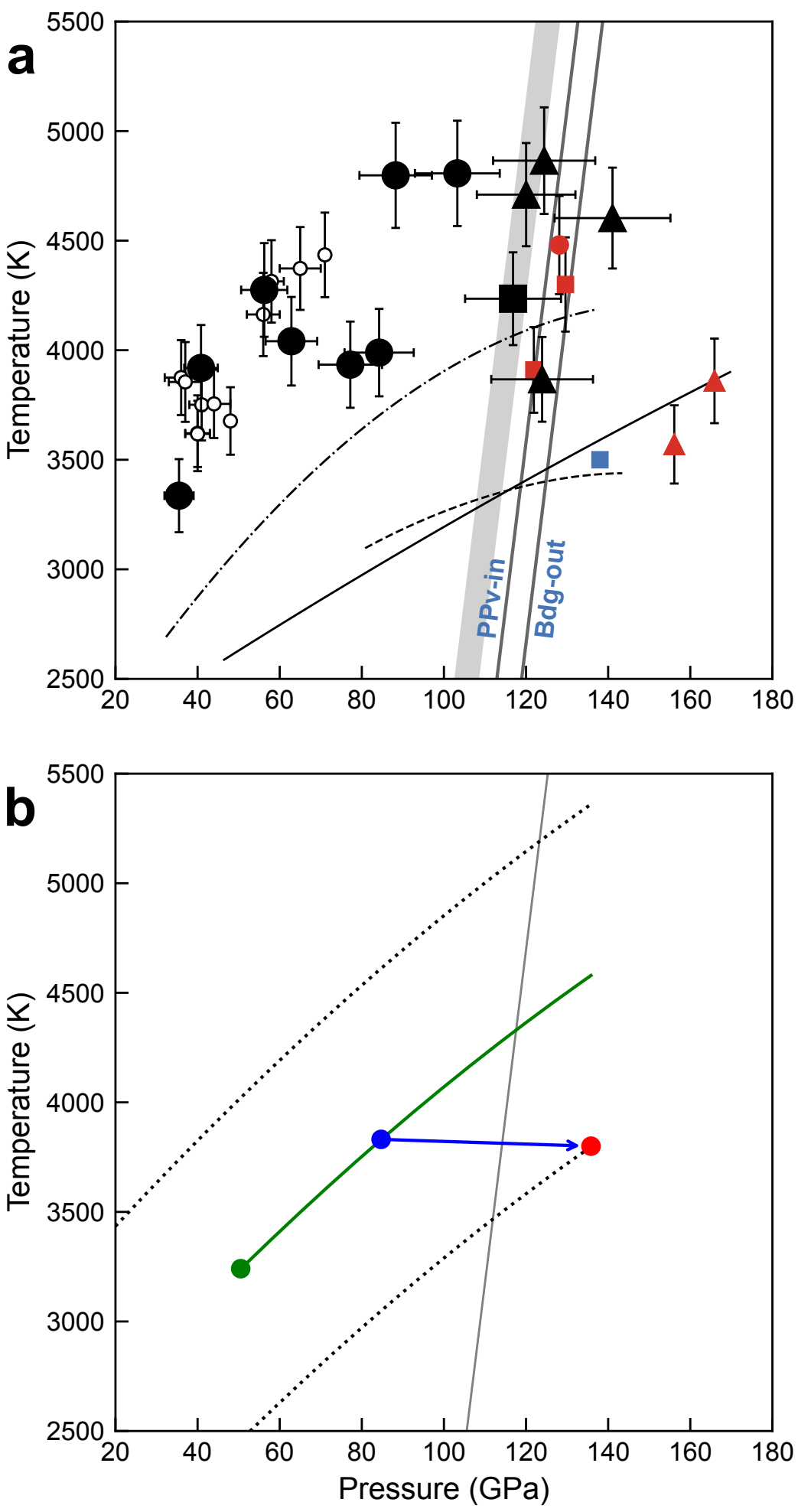

**Extended Data Fig. 1 | Pressure and temperature conditions of the present melting experiments and for the Bdg/PPv crystallization in the BMO evolution modelling. a**, Melt coexisted with Bdg (circles), PPv (triangles) or both (squares) in our experiments (large black filled symbols), earlier hydrous experiments[10] (small open circles), anhydrous experiments[32] (small blue square) and those[50] possibly containing ~1,500 ppm $H_2O$ in their starting material (small red symbols). The PPv-in and Bdg-out curves are from ref. 50. PPv appeared in the present experiments at pressures slightly lower than those reported by Kuwayama *et al.*[50]. The grey band shows a tentative Bdg-PPv boundary in a hydrous BMO based on the coexistence of Bdg, PPv and melt in run #9 (black square). Thin solid, broken and dashed-dotted curves represent the solidus temperatures of pyrolite reported by refs. 60, 74 and 32, respectively. **b**, The upper dotted curve represents the liquidus of a pyrolitic mantle[32]. The temperature of the present-day ULVZs may be 3,800 K (red circle, equivalent to the CMB temperature), which could approximate the temperature profile (lower dotted curve), at which fractional crystallization terminates. We suppose that the 50% batch crystallization *P-T* profile (green curve) lies in the middle between the liquidus and crystallization-terminate temperature curves. Subsequent fractional crystallization takes place along the blue profile. Grey line represents the Bdg-PPv phase transition boundary in a hydrous BMO according to **a**. See Methods for modelling details.

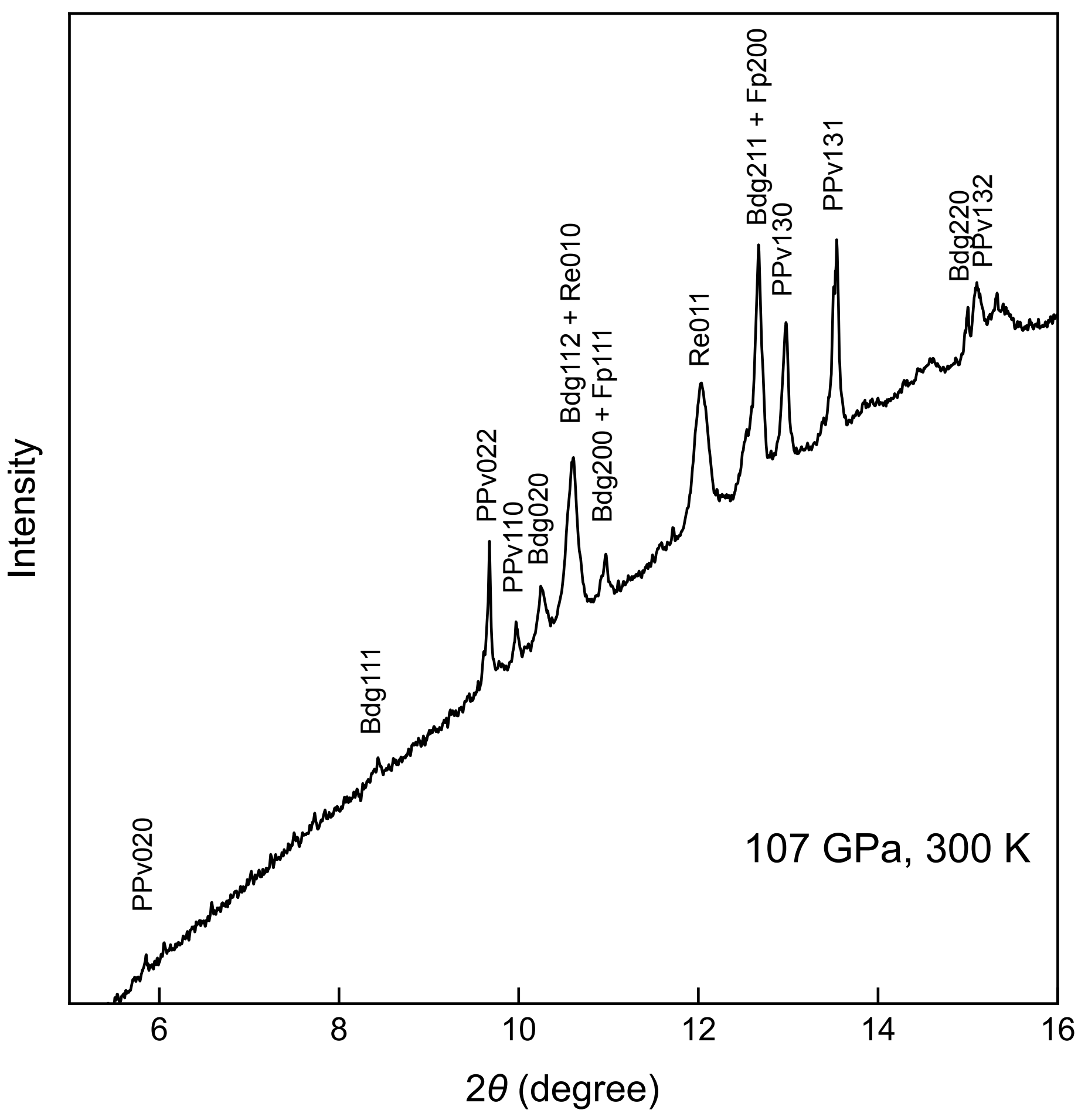

**Extended Data Fig. 2 | XRD observation of the liquidus phases of Bdg and PPv coexisting with melt collected after quenching temperature to 300 K in run #9.** This XRD pattern was collected 9 μm away from the centre of a laser-heated hot spot.

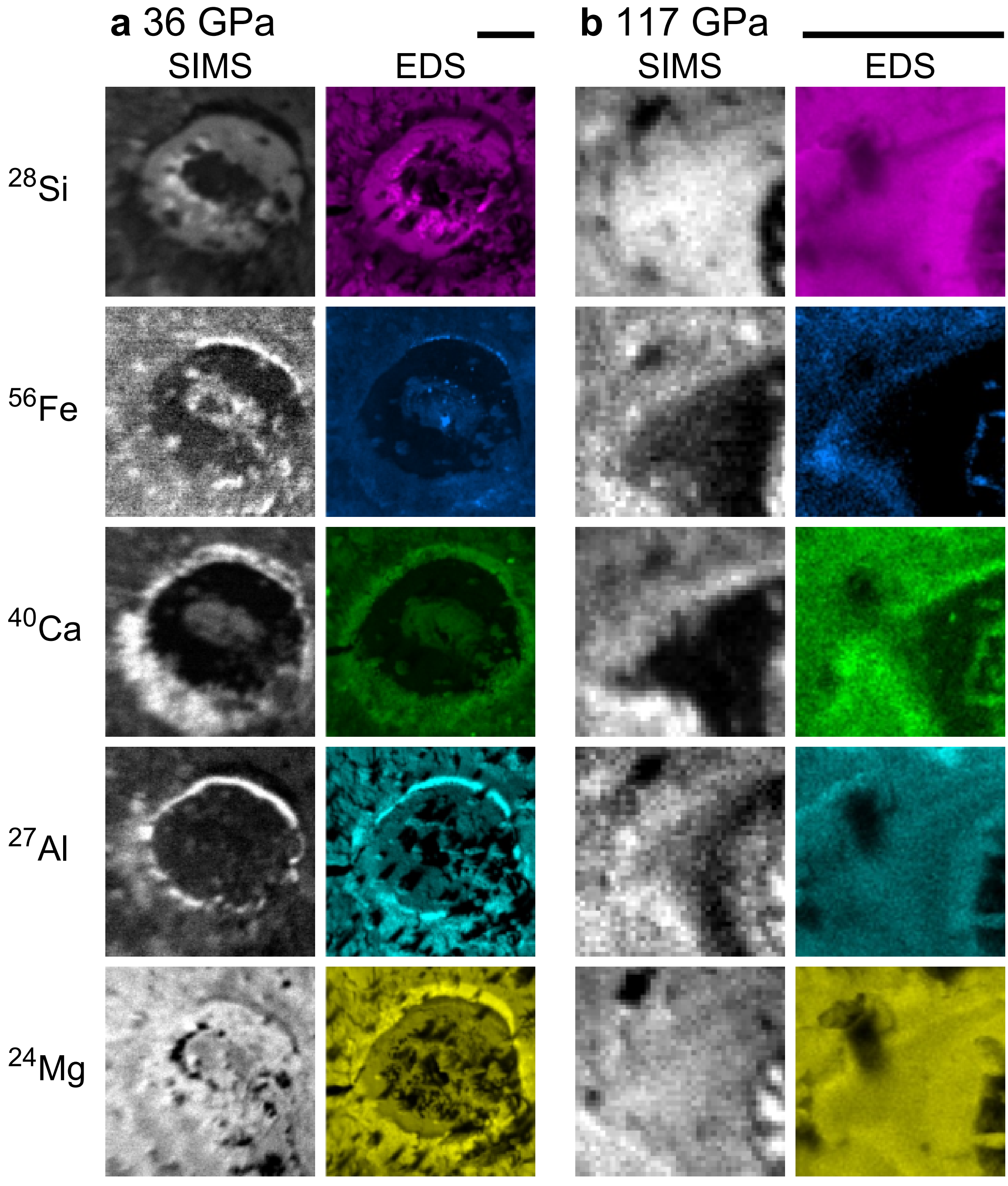

**Extended Data Fig. 3 | Cryo-SIMS and EDS X-ray maps for Si, Fe, Ca, Al and Mg.** See Fig. 1 for the $^{1}$H and $^{2}$D maps. All images in linear scale. Scale bars, 20 μm.

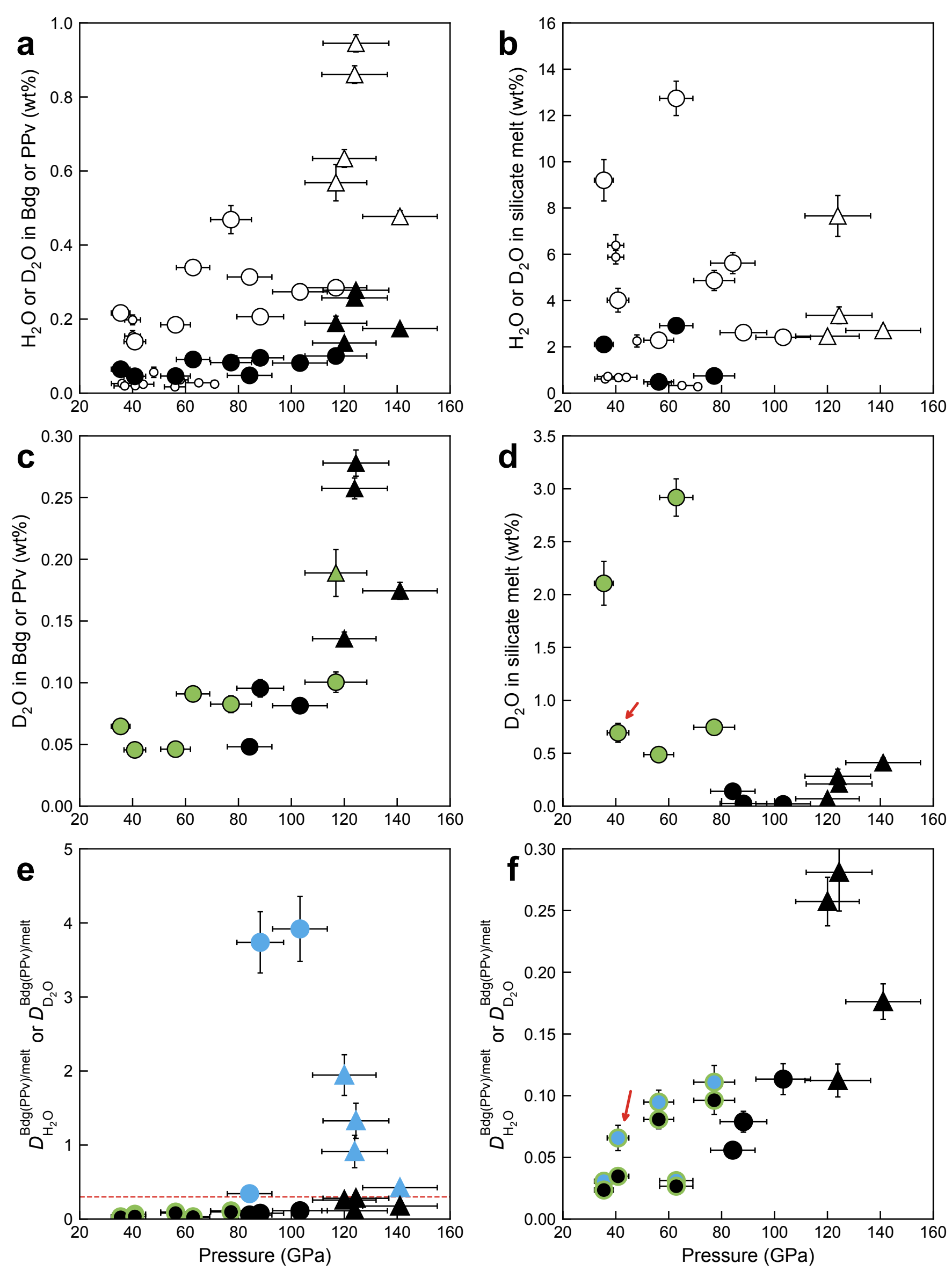

**Extended Data Fig. 4 | SIMS analyses of $H_2O$ and $D_2O$ concentrations in coexisting Bdg/PPv crystals and melt and resulting crystals/melt partition coefficients.** Circles, Bdg; triangles, PPv. **a,b,** $H_2O$ (open) and $D_2O$ concentrations (closed) in Bdg/PPv crystals (**a**) and melt (**b**). Small symbols are from ref. 10. **c,d,** Close-up for $D_2O$. The analyses of melts include both samples decompressed/recovered at cryogenic (green) and ambient temperatures (black) (**d**). The latter samples are distinguished from the former by their low concentrations and exceedingly high Bdg(PPv)/melt $D_{D2O}$ in **e**, suggesting the loss of $D_2O$ from quenched melt upon decompression under 300 K, and therefore not included in **b**. **e,f,** Comparison of Bdg(PPv)/melt partition coefficients between $H_2O$ (black) and $D_2O$ (blue). All runs (**e**) and enlarged (**f**) below the red dashed line in **e**. Samples decompressed/recovered at cryogenic temperatures are framed by green. In run #2 marked by a red arrow (**d,f**), deuterium was likely to be partially lost from melt due to temperature increase during SIMS measurements (see Methods for details), causing apparently remarkably higher $D(D_2O)$ than $D(H_2O)$ (**f**).

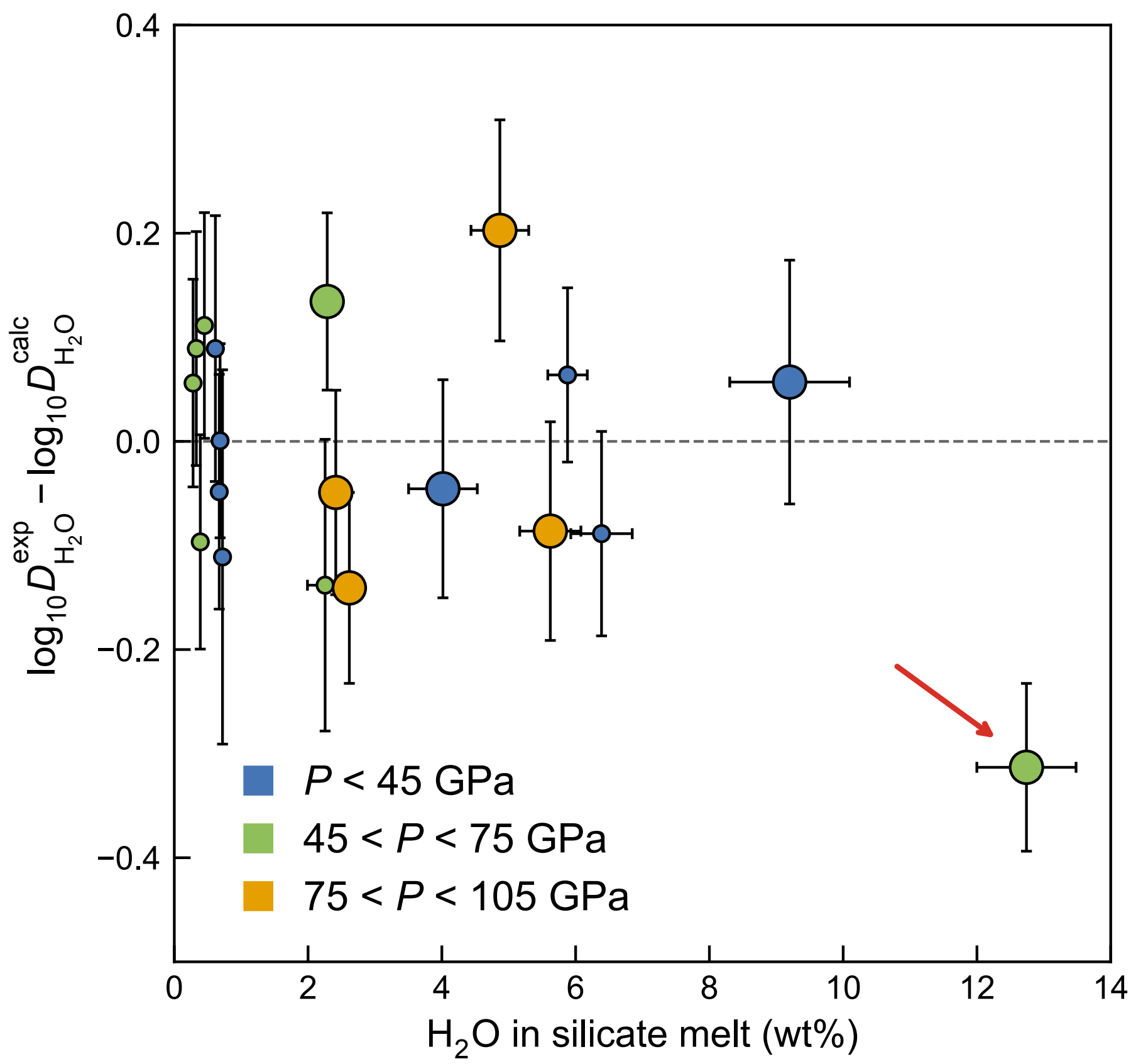

**Extended Data Fig. 5 | Variations in the Bdg/melt partition coefficient of $H_2O$ with changing water concentration in melt.** The difference in $D_{H2O}$ between measured and calculated values from the fitting results (Eq. 1) is shown on the vertical axis. Large and small symbols are from this study and ref. 10, respectively. A strong reduction in $D_{H2O}$ (shown by red arrow) suggests that Bdg was saturated with $H_2O$ in run #4, while undersaturated in other runs.

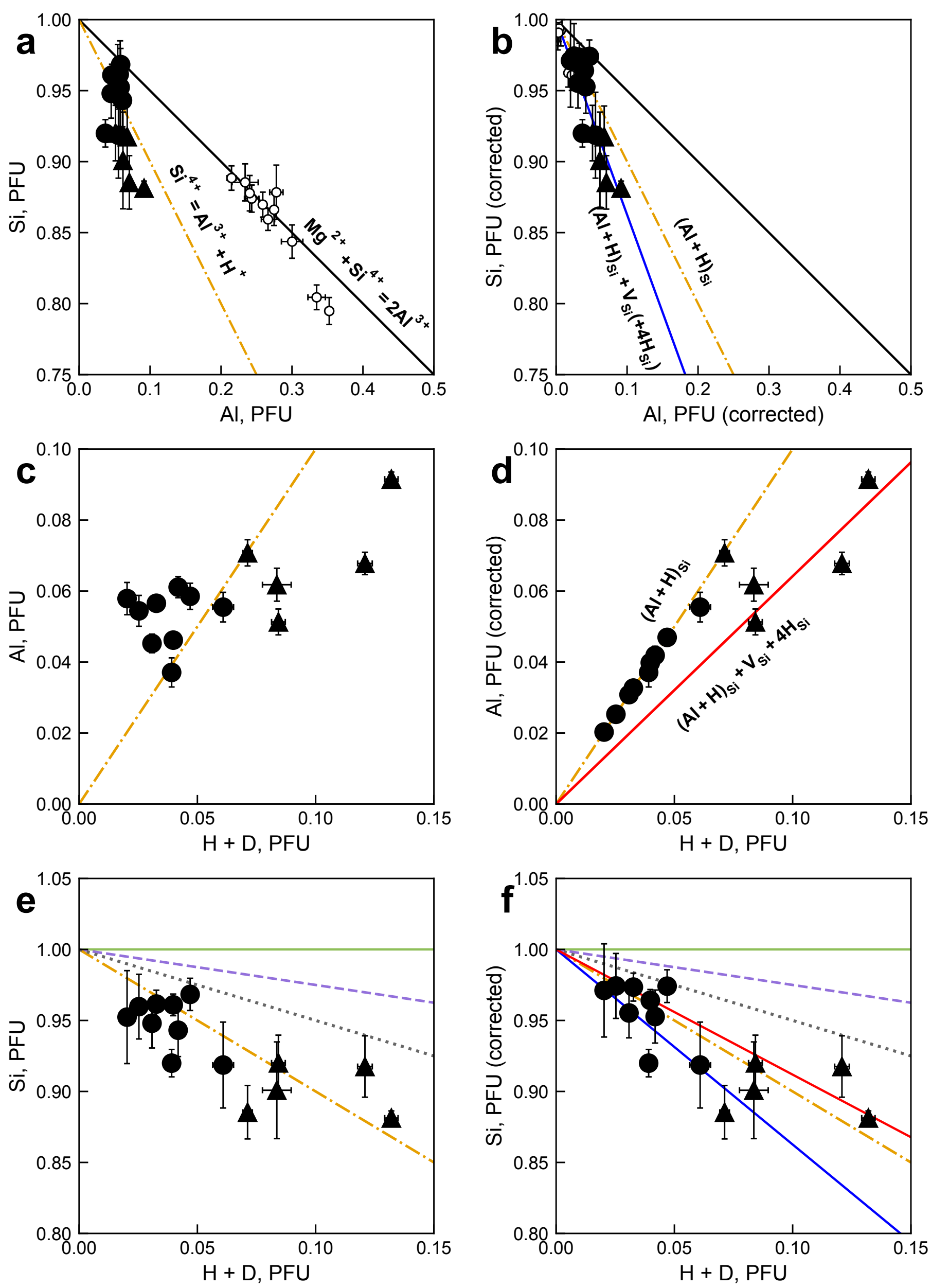

**Extended Data Fig. 6 | Variations in the Si, Al and H contents in Bdg (circles) and PPv (triangles).** Green, purple, grey and orange lines denote co-variations formed by the $Mg^{2+} = 2H^+$ (including $D^+$ for each substitution), $Si^{4+} = 4H^+$, $Si^{4+} = Mg^{2+} + 2H^+$ and $Si^{4+} = Al^{3+} + H^+$ substitutions, respectively. Raw analyses per formula unit (Extended Data Table 1) (**a, c, e**) and after correction for the $Mg^{2+} + Si^{4+} = 2Al^{3+}$ substitution (considering Al is incorporated by the $Si^{4+} = Al^{3+} + H^+$ and $Mg^{2+} + Si^{4+} = 2Al^{3+}$ substitutions) (**b, d, f**). $Si^{4+}$ was mainly substituted by $Al^{3+} + H^+$ in this study (filled symbols), while the $Mg^{2+} + Si^{4+} = 2Al^{3+}$ substitution was predominant in earlier experiments[10] (open symbols) (**a**). While such $Si^{4+} = Al^{3+} + H^+$ substitution explains water incorporation into Bdg, it does not fully account for that into PPv (**c, d**). The depletion in Si in PPv indicates the formation of Si vacancies (blue line in **b**, **f**), which may be partially occupied by 4H (red line in **d**, **f**).

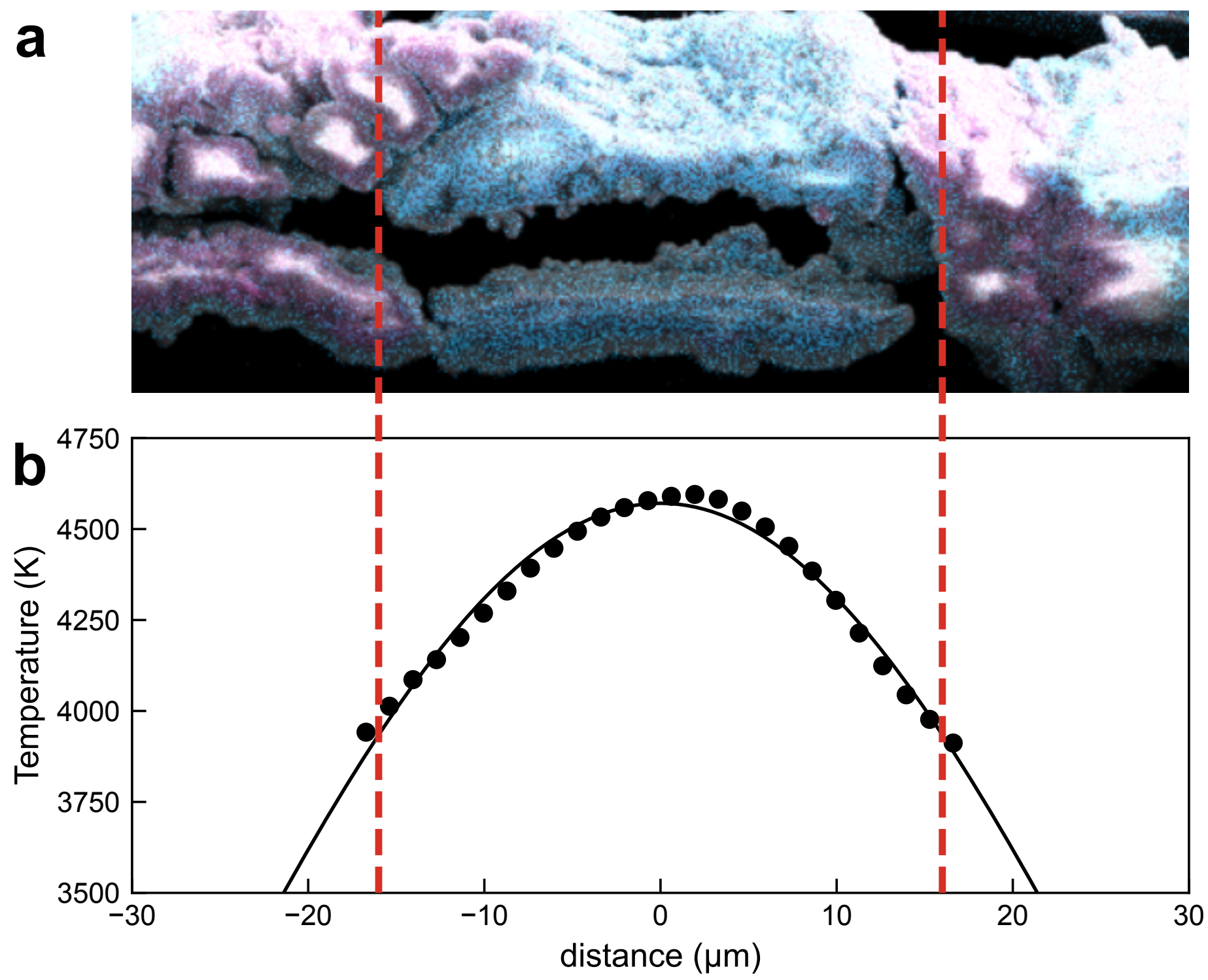

**Extended Data Fig. 7 | Sample cross section and temperature profile for run #2 at 41 GPa.** This cross section was prepared after SIMS measurements. **a,** A map combining back-scattered electron image and EDS X-ray maps for Si (magenta) and Ca (cyan). **b,** Corresponding temperature profile across the centre of a laser-heated spot. Red dashed lines mark the boundaries between quenched melt and the liquidus phase (Bdg), showing the temperature was 3,920 K at the boundary (the temperature at the melt-Bdg boundary should have been identical for both sides).

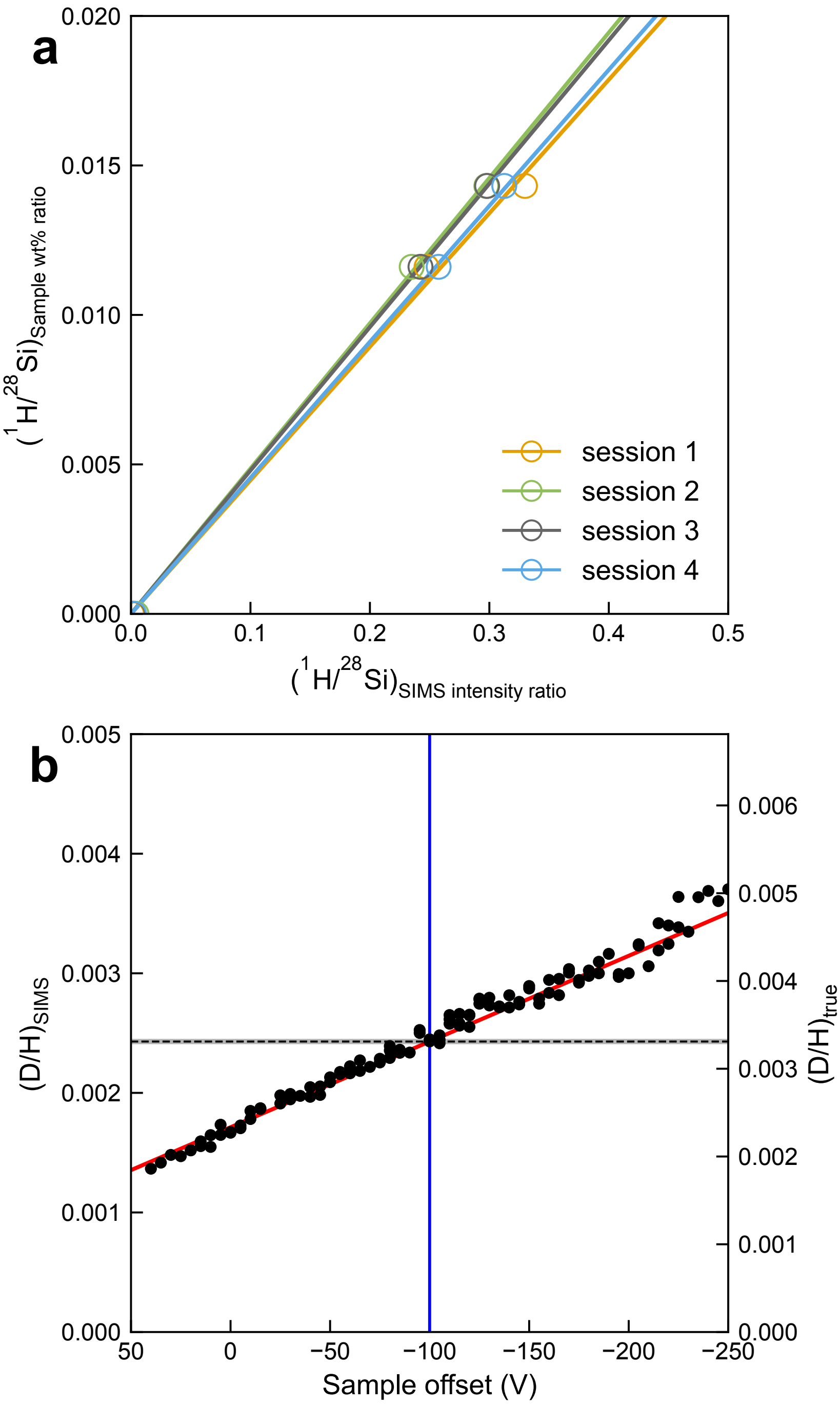

**Extended Data Fig. 8 | Calibration curves for the SIMS analyses of $^{1}H/^{28}Si$ and D/H ratio. a,** The $^{1}H/^{28}Si$ ratios were measured for three standard glasses (one anhydrous and two hydrous) in each analytical session. **b,** The SIMS D/H count ratio changes with varying the sample offset voltage. The red line shows the regression. The offset of –100 V employed for the measurements in this study is highlighted by the blue line. The D/H IMF is obtained for measured D/H count to be consistent with the true D/H value (grey line with an error band) at the sample offset voltage of –100 V.

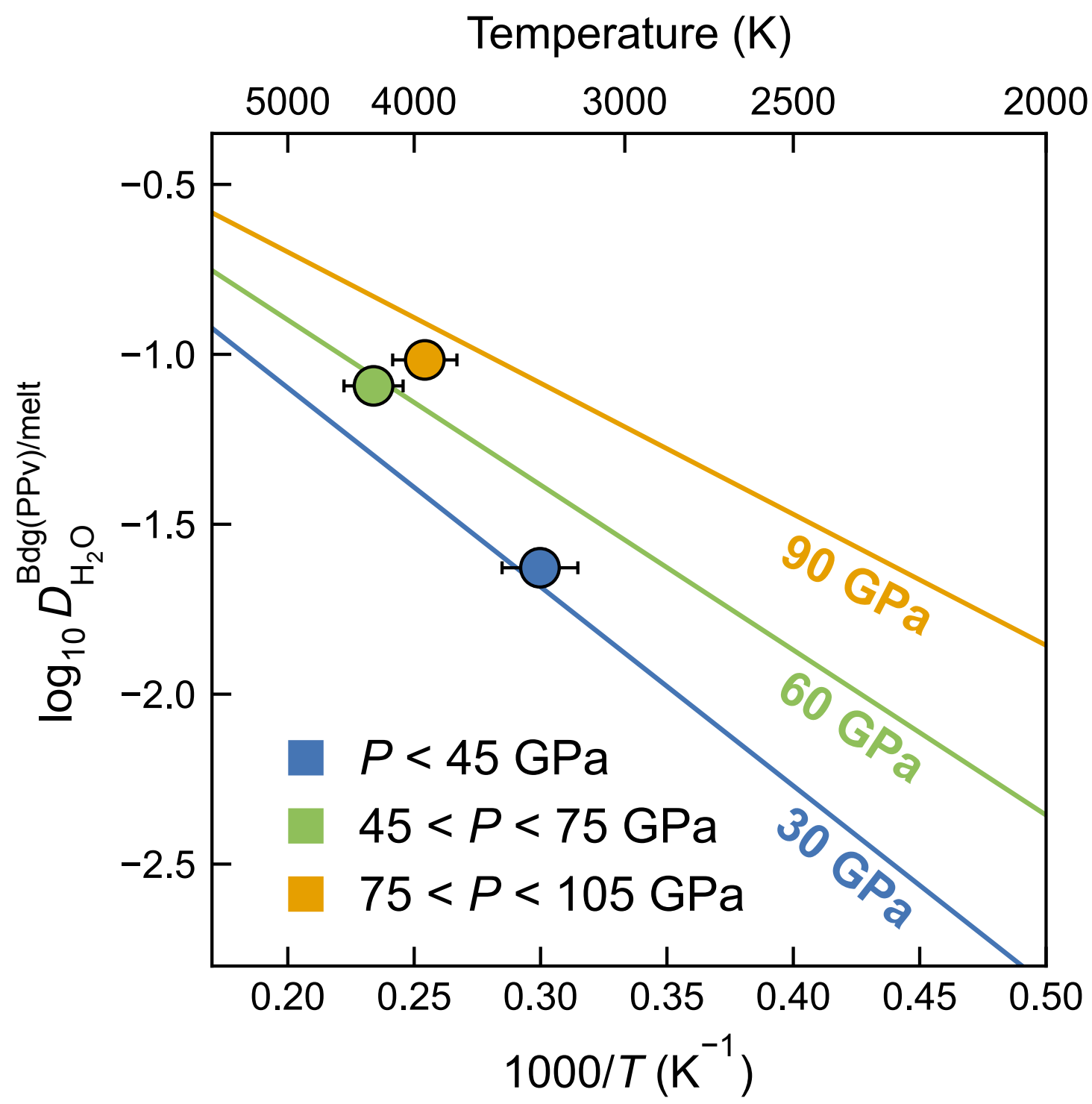

**Extended Data Fig. 9 | Pressure and temperature dependences of $D_{H2O}$(Bdg/melt) based on the data only from runs with cryogenic sample recovery.** We obtain $\log_{10} D_{H2O}$(Bdg/melt) = 0.07 – 6.9 × $10^3/T$ + 33.3 × $P/T$, each parameter consistent within 1σ uncertainty with that in Eq. 1 (see the main text).